\documentstyle[12pt,equation]{article}
\begin{document}
\newcommand{\tcr}{T_{cr}}
\newcommand{\df}{\delta \phi}
\newcommand{\dkl}{\delta \kappa_{\Lambda}}
\newcommand{\lx}{\lambda}
\newcommand{\Lx}{\Lambda}
\newcommand{\ex}{\epsilon}
\newcommand{\db}{{\bar{\delta}}}
\newcommand{\lb}{{\bar{\lambda}}}
\newcommand{\lt}{\tilde{\lambda}}
\newcommand{\lr}{{\lambda}_R}
\newcommand{\lbr}{{\bar{\lambda}}_R}
\newcommand{\lk}{{\lambda}(k)}
\newcommand{\lbk}{{\bar{\lambda}}(k)}
\newcommand{\ltk}{\tilde{\lambda}(k)}
\newcommand{\mx}{{m}^2}
\newcommand{\mb}{{\bar{m}}^2}
\newcommand{\mt}{\tilde{m}^2}
\newcommand{\mr}{{m}^2_R}
\newcommand{\mk}{{m}^2(k)}
\newcommand{\Pt}{\tilde P}
\newcommand{\Mt}{\tilde M}
\newcommand{\Qt}{\tilde Q}
\newcommand{\Nt}{\tilde N}
\newcommand{\rhb}{\bar{\rho}}
\newcommand{\rht}{\tilde{\rho}}
\newcommand{\rhz}{\rho_0}
\newcommand{\yz}{y_0}
\newcommand{\rhzk}{\rho_0(k)}
\newcommand{\kx}{\kappa}
\newcommand{\kt}{\tilde{\kappa}}
\newcommand{\kk}{\kappa(k)}
\newcommand{\ktk}{\tilde{\kappa}(k)}
\newcommand{\Gammat}{\tilde{\Gamma}}
\newcommand{\Lt}{\tilde{L}}
\newcommand{\Zt}{\tilde{Z}}
\newcommand{\zt}{\tilde{z}}
\newcommand{\zh}{\hat{z}}
\newcommand{\uh}{\hat{u}}
\newcommand{\Mh}{\hat{M}}
\newcommand{\wt}{\tilde{w}}
\newcommand{\etat}{\tilde{\eta}}
\newcommand{\Gammak}{\Gamma_k}
\newcommand{\be}{\begin{equation}}
\newcommand{\ee}{\end{equation}}
\newcommand{\een}{\end{subequations}}
\newcommand{\ben}{\begin{subequations}}
\newcommand{\beq}{\begin{eqalignno}}
\newcommand{\eeq}{\end{eqalignno}}
\renewcommand{\thefootnote}{\fnsymbol{footnote} }
\pagestyle{empty}
\noindent
DESY 93-094 \\
HD-THEP-93-28 \\
July 1993
\vspace{3cm}
\begin{center}
{{ \Large  \bf
Critical Exponents  \\
from the Effective Average Action
}}\\
\vspace{10mm}
N. Tetradis \\
{\em Deutsches Elektronen-Synchrotron DESY, Gruppe Theorie,\\
Notkestr. 85, 22603 Hamburg} \\
\vspace{5mm}
and \\
\vspace{5mm}
C. Wetterich \\
{\em
Institut f\"ur Theoretische Physik, Universit\"at Heidelberg,\\
Philosophenweg 16, 69120 Heidelberg}\\
\end{center}

\setlength{\baselineskip}{20pt}
\setlength{\textwidth}{13cm}

\vspace{3.cm}
\begin{abstract}
{
We compute the critical behaviour of three-dimensional scalar theories
using a new exact non-perturbative evolution equation. Our values for
the critical exponents agree well with previous precision estimates.
}
\end{abstract}
\clearpage
\setlength{\baselineskip}{15pt}
\setlength{\textwidth}{16cm}
\pagestyle{plain}
\setcounter{page}{1}

\newpage

\setcounter{equation}{0}
\renewcommand{\theequation}{{\bf 1.}\arabic{equation}}

\section*{1.
Introduction}

A new exact non-perturbative evolution equation for the scale
dependence of an effective action has been proposed recently
\cite{exact}. The effective average action $\Gamma_k$ obtains
by integrating out only the quantum fluctuations with momenta
$ q^2 \geq k^2$. Then the usual effective action (the generating
functional of the 1PI Green functions)
is recovered in the limit $k \rightarrow 0$ where all quantum
fluctuations are included.
The effective average action can also be viewed as the action
for averages of fields, similar in spirit to the block spin
action \cite{kadanoff,wilson} in lattice theories.
The dependence of the effective average action on the scale $k$ is
described by the exact evolution equation. This equation has the
form of a one-loop expression, differentiated with respect to $k$.
The non-perturbative content, which makes this equation exact, consists
of using the exact inverse propagator, as given by the second functional
derivative of the effective average action $\Gamma_k$, instead of the
inverse propagator. Furthermore, an infrared cutoff suppressing
the fluctuations with $q^2 \ll k^2$ is added to the inverse propagator.
In consequence, one obtains an equation where the scale dependence of
$\Gamma_k$ is expressed in terms of its second functional
derivative with respect to the fields
($t=\ln k$)
\be
\frac{\partial}{\partial t} \Gammak
= \frac{1}{2} {\rm Tr} \bigl\{ (\Gammak^{(2)} + R_k)^{-1}
\frac{\partial}{\partial t} R_k \bigr\}.
 \label{oneone} \ee
(Here $R_k$ is the effective infrared cutoff - details will be given in
the next section.)
The momentum integration implied by the trace is both ultraviolet and
infrared finite in arbitrary dimensions. This makes our non-perturbative
equation
suitable for dealing with theories which are plagued by infrared
problems in perturbation theory.
(Scalar theories in less than four dimensions or at non-vanishing
temperature near a second order phase transition and non-abelian gauge theories
figure among the examples.)
\par
The most important new feature seems to us not the exactness of
eq. (\ref{oneone}), but its simple and intuitive form which makes it
suitable for approximate solutions.
Exact equations have been derived earlier, as for example
the Schwinger-Dyson equations \cite{schwinger}
or the ``exact renormalization group equations'' \cite{wilson,polchinski}.
(There exists actually a formal relation between the latter and
eq. (\ref{oneone}) \cite{private} - \cite{mmore}.
There should also be a formal relation to the Schwinger-Dyson equations -
after all, exact relations describing the same Green functions should all
be equivalent.)
The evolution equation (\ref{oneone}) is a functional differential equation.
It can be rewritten as an infinite system of ordinary coupled non-linear
differential equations for the infinite number of couplings
appearing in the most general form of $\Gamma_k$ consistent with the
symmetries \cite{mmore}.
(For example, the general potential of a scalar field theory contains
terms $\sim \phi^6$, $\sim \phi^8$ etc.) An exact solution of this infinite
system seems almost impossible for an interacting theory. The r.h.s.
of eq. (\ref{oneone}) involves the exact propagator which is as difficult to
compute as $\Gamma_k$ itself. The main point is that for many situations of
interest the propagator is approximately known. Then one can employ the
strategy to describe the propagator in terms of a finite number of parameters
$\alpha_i$ which may depend on $k$. Typically, such parameters are masses, wave
function renormalization constants etc. This approximation corresponds to an
ansatz for the average action $\Gamma_k$ in terms of a finite number of
undetermined functions $\alpha_i(k)$. The evolution equation
(\ref{oneone}) is now truncated and turned into a finite number of
differential equations
\be
\frac{d \alpha_i}{d t} = \beta_i(\alpha_j).
\label{onetwo} \ee
Such a finite system can be solved numerically or sometimes
analytically. The sucess of our method ultimately depends on a good
guess about the general form of the exact propagator, which
in turn depends on the proper choice of degrees of freedom. (Sometimes
it may be useful to include composite fields \cite{ellwanger}.)
Fortunately, eq. (\ref{oneone}) has a form very similar to the
one-loop expression in standard perturbation theory.
(It reproduces perturbation theory in its range of applicability.
The loop expansion can be recovered from eq. (\ref{oneone})
by a systematic expansion of the solution in terms of the small
coupling.)
This allows to build on the experience gained from the perturbative
renormalization group analysis \cite{schwinger}, \cite{calsym} -
\cite{parisi} and provides useful checks for situations of overlap.
Nevertheless, our equation is genuinely non-perturbative.
Scalar theories at the critical temperature or in two and three
dimensions are not accessible to perturbation theory and are
described by our method with success \cite{christof2,transition}.
\par
For a scalar theory the inverse propagator is characterized by a
mass matrix and a generalized kinetic term $\sim Z q^2$. The mass
matrix is directly related to derivatives of the scalar potential
$U_k$ which describes the non-derivative part of
$\Gamma_k$. In general $Z$ depends on fields and momenta and its
detailed structure may be quite complicated.
Nevertheless, this quantity plays essentially the role of a field and
momentum dependent wave function renormalization.
As a result its dependence on the
various mass scales of the model is governed
by the anomalous dimension.
In particular, one expects a weak dependence on fields, momenta and the scale
$k$ if the anomalous dimension $\eta$ is small. This is typically the case
in three- and four-dimensional scalar theories. We will exploit this fact
in the present paper.
\par
The aim of this paper is a precision test of our method in a context which
is not accessible to perturbation theory. We consider the three-dimensional
$O(N)$-symmetric scalar theory in the phase with spontaneous symmetry
breaking. The presence of the Goldstone bosons leads to severe
infrared divergences in perturbation theory and excludes a
perturbative treatment in three dimensions. On the other hand, the model
exhibits a second order phase transition. The critical exponents at the
phase transitions have been computed by various other methods
\cite{zinn,parisi} with high precision. They serve as a valuable test.
We shall see that even a truncation to only a few parameters $\alpha_i$
gives good quantitative results for the critical exponents, typically with
an accuracy at the percent level. In view of the fact that the
momentum integration in eq. (\ref{oneone}) is the same as for a
one-loop calculation, and that no small coupling characterizes the
interactions at the phase transition, this result seems quite remarkable.
\par
Since all our computations are done directly in three dimensions, we are
not limited to the vicinity of the phase transition. In particular,
we can employ our method in the case of the four-dimensional
scalar theory at non-vanishing temperature and follow the transition
from the effective four-dimensional behaviour for $k \gg T$ to an
effective three-dimensional behaviour for $k \ll T$ \cite{transition}.
The second order character of the high temperature transition
in $O(N)$-symmetric scalar theories has been established previously
\cite{transition}, together with a crude estimate of the critical
exponents. The present work can be seen as a refinement of
\cite{transition} near the critical temperature where the
three-dimensional behaviour dominates. A more accurate quantitative
description of the temperature dependence of masses, expectation values
and couplings away from the critical temperature, or the study of the
full temperature dependent scalar potential will require to redo the
computations of \cite{transition}, including the additional couplings
discussed in the present paper.
\par
This paper is organized as follows: We present the general formalism
in section 2 and derive in section 3 an exact evolution equation for the
scalar potential formulated in terms of dimensionless couplings.
No explicit mass scale appears in this equation, which is
therefore well suited for a description of the scale independent physics
at the second order phase transition. In section 4 we approximate this
equation with a uniform (field and momentum independent) wave function
renormalization. This approximation becomes exact in the large $N$ limit.
In section 5 we solve the exact evolution equation in the large $N$ limit
and establish consistency with earlier results \cite{jain,largen}. In
section 6 we discuss the general form of the effective scalar
potential at the phase transition for arbitrary values of $N$ by
taking the limit $k \rightarrow 0$ for fixed values of the fields. In
section 7 we exploit the knowledge that the anomalous dimension is
small and sketch a truncation adapted to this property. The anomalous
dimension is finally computed in section 8.
Our results for the critical exponents are presented in section 9 and
section 10 gives our conclusions.

\setcounter{equation}{0}
\renewcommand{\theequation}{{\bf 2.}\arabic{equation}}

\section*{2.
Exact evolution equation for the effective average action}

The scale dependent effective action $\Gamma_k$,
for a theory described by a classical action $S$,
results from the effective integration
of degrees of freedom with characteristic momenta larger than a given
infrared cutoff $k$. The dependence of the effective action
on the scale $k$ is described by an exact evolution equation.
In this section we summarize the formalism
for the implementation of the above ideas in the case of a
$O(N)$-symmetric theory of real scalar fields.
For a detaided discussion, proof of the exactness of the evolution
equation and modification of the conventions for the discussion
of complex fields,
we refer the reader to ref. \cite{exact}. For our conventions see also
the appendix A of ref. \cite{christof1}.
\par
We consider a theory of $N$ real scalar fields $\chi^a$, in
$d$ dimensions, with an
$O(N)$-symmetric action $S[\chi]$.
We specify the action together with some ultraviolet
cutoff $\Lambda$, so that the theory is properly
regulated.
We add to the kinetic term a piece
\be
\Delta S = \frac{1}{2} \int d^d q
R_k(q) \chi^*_a(q) \chi^a(q).
\label{twoone} \ee
The function $R_k$  is employed in
order to prevent the propagation of modes
with characteristic momenta $q^2 < k^2$.
This can be achieved, for example,
by the choice
\be
R_k(q) = \frac{Z_k q^2 f^2_k(q)}{1 - f^2_k(q)},
\label{twotwo} \ee
with
\be
f^2_k(q) = \exp \left( - \frac{q^2}{k^2} \right).
\label{twothree} \ee
As a result the inverse propagator
for the action $S + \Delta S$ has a minimum $\sim k^2$.
The modes with $q^2 \gg k^2$ are unaffected,
while the
low frequency modes with $q^2 \ll k^2$ are cut off:
\be
\lim_{q^2 \rightarrow 0} R_k = Z_k k^2.
\label{twofour} \ee
The quantity $Z_k$ in eq. (\ref{twotwo})
is an appropriate wave function renormalization whose
precise definition will be given below.
We emphasize at this point that the above form of $R_k$ is not unique
and many alternative choices are possible.
We subsequently introduce sources and
define the generating functional for the connected Green functions
for the action $S + \Delta S$. Through a Legendre
transformation we obtain the
generating functional for the 1PI Green functions
${\tilde \Gamma}_k[\phi^a]$, where $\phi^a$ is the expectation value of the
field $\chi^a$ in the presence of sources.
The use of the modified propagator for the calculation of
${\tilde \Gamma}_k$ results in the effective integration of only the
fluctuations with $q^2 \geq
k^2$. Finally, the scale dependent effective action is
obtained by removing the infrared cutoff
\be
\Gamma_k[\phi^a] = {\tilde \Gamma}_k[\phi^a] -
\frac{1}{2} \int d^d q
R_k(q) \phi^*_{a}(q) \phi^a(q).
\label{twofive} \ee
For $k$ equal to the ultraviolet cutoff $\Lambda$, $\Gammak$ becomes
equal
to the classical action $S$ (no effective integration of modes takes
place), while for $k \rightarrow 0$ it tends towards the effective action
$\Gamma$ corresponding to $S$ (all the modes are included).
The interpolation of $\Gammak$ between the classical and the
effective action makes it a very useful field theoretical tool.
The means for practical calculations is provided by an exact
evolution equation which describes the response of the
scale dependent effective action to variations of the infrared cutoff
($t=\ln(k/\Lambda)$) \cite{exact}:
\be
\frac{\partial}{\partial t} \Gammak[\phi]
= \frac{1}{2} {\rm Tr} \bigl\{ (\Gammak^{(2)}[\phi] + R_k)^{-1}
\frac{\partial}{\partial t} R_k \bigr\}.
 \label{twosix} \ee
Here $\Gammak^{(2)}$ is the second functional derivative of the scale dependent
effective action with respect to $\phi^a$. For
real fields in momentum space
\footnote{
In order to avoid excessive formalism, we shall use the same notation
for functions or operators in position space and their Fourier transforms,
defined with the convention
$\phi(q)=(2 \pi)^{-\frac{d}{2}}\int d^dx \exp(iq_{\mu}x^{\mu}) \phi(x)$.
}, it reads
\be
(\Gammak^{(2)})^a_b(q,q') =
\frac{\delta^2 \Gammak}{\delta \phi^*_a(q) \delta \phi^b(q')},
\label{twoseven} \ee
with
\be
\phi^a(-q)=\phi^*_a(q).
\label{twoeight} \ee
Note also that, although not explicitly
indicated for simplicity, $R_k$ in eq. (\ref{twosix}) stands for
\be
(R_k)^a_b(q,q') = R_k(q) \delta^a_b \delta(q-q').
\label{twonine} \ee
In momentum space,
for a system whose volume $\Omega$ is taken to infinity at the end,
the trace reads
\be
{\rm Tr} = \Omega \sum_{a} \int \frac{d^dq}{(2 \pi)^d},
\label{twoten} \ee
and it is assumed to act on the diagonal part
of the generalized matrix.
The exact evolution equation gives the response of the scale dependent
effective
action $\Gammak$ to variations of the scale $k$, through a one-loop expression
involving the exact inverse propagator $\Gamma^{(2)}_k$ together with an
infrared cutoff provided by $R_k$.
It has a simple graphical representation (fig. 1). Our non-perturbative
exact evolution equation can be
viewed as a partial differential equation for
the infinitely many variables $t$ and $\phi^a(q)$.
Its usefulness depends on the existence of
appropriate truncations which permit its solution.
\par
Before presenting the formalism which leads to approximate solutions of
eq. (\ref{twosix}), we briefly comment on the relation of the scale
dependent effective
action to the average action, which has been used for practical calculations in
the past. The average action was introduced in ref. \cite{christof1,christof2}
in order to describe the dynamics of averages of fields over volumes
$\sim k^{-d}$. The implementation of an infrared cutoff $k$ was
naturally incorporated through
the averaging procedure. The dependence of the average action on $k$
was computed through
an one-loop renormalization group equation which corresponds to
a truncation of
eq. (\ref{twosix}). It was shown in ref. \cite{more} that
a modification of the averaging procedure leads to an improved
average action which is identical to the scale dependent effective action
up to an explicitely known ultraviolet regulator in the improved average
action. For this reason we call $\Gamma_k$ the effective average action.
Within the approximations used so far there is no difference for low
momentum quantities between the originally proposed average action
\cite{christof1,christof2}, the improved average action \cite{more} and
the effective average action \cite{exact}.
As a result, all previous calculations can be viewed as
approximate solutions of eq. (\ref{twosix}) with some appropriate truncation.
More specifically, these previous studies have described correctly the
phase structure of the two- and three- dimensional
scalar theories (including the
Kosterlitz-Thouless phase transition) \cite{christof2}, the high
temperature phase transition of the four-dimensional theory
(with a reliable description of the effectively three-dimensional
critical behaviour and a rough
determination of critical exponents) \cite{transition, largen}, and the
approach to convexity for the effective potential
in the phase with spontaneous symmetry breaking \cite{convex}.
\par
As we have already mentioned, for the solution of eq. (\ref{twosix})
one has to develop an efficient truncation scheme.
We consider
an effective action of the form
\be
\Gammak =
\int d^dx \bigl\{ U_k(\rho)
+ \frac{1}{2} \partial^{\mu} \phi_a Z_k(\rho,-\partial^{\nu}\partial_{\nu})
\partial_{\mu} \phi^a
+ \frac{1}{4} \partial^{\mu} \rho Y_k(\rho,-\partial^{\nu}\partial_{\nu})
\partial_{\mu} \rho
\bigr\},
\label{twoeleven} \ee
where $\rho=\frac{1}{2} \phi_a \phi^a$
and appropriate normal ordering is assumed for the derivative operators
\cite{exact}.
In order to turn the evolution equation for the
effective average action into equations for
$U_k$, $Z_k$ and $Y_k$, we have to evaluate the trace in eq. (\ref{twosix})
for properly chosen
background field configurations.
For the evolution equation for $U_k$ we have to expand around a constant
field configuration. This calculation is carried out in section 3 and
the study of the resulting equation occupies sections 4-7.
The evolution of the wave function renormalizations, which leads to the
determination of the anomalous dimensions, requires an expansion around
a background with a small momentum dependence. This calculation is
presented in section 8 and appendix B.
The evolution equations for $U_k$, $Z_k$ and $Y_k$ are partial differential
equations for independent variables $t$ and $\rho$. In most cases
their study is
possible when they are
turned into an (infinite) set of coupled ordinary differential equations for
independent variable $t$.
This is achieved by Taylor expanding
$U_k$, $Z_k$ and $Y_k$ around some value
of $\rho$. Since in this work we are interested in the vacuum structure of the
theory, we shall use an expansion around
the $k$ dependent minimum $\rhzk$ of $U_k$.
In the
limit $k \rightarrow 0$ this minimum
specifies the vacuum of the theory, and
$U=U_0$, $Z=Z_0$ and $Y=Y_0$
and their $\rho$-derivatives at the minimum give the
renormalized masses, couplings and wave function renormalizations.
\par
In a first approximation
we can neglect the $q^2$ dependence of
$Z_k(\rho,q^2)$.
We can now define the quantity $Z_k$ appearing in
eq. (\ref{twotwo})
as
\be
Z_k = Z_k(\rhzk).
\label{twotwelve} \ee
For studies which
concentrate on the minimum of $U_k$,
the above definition permits the combination of
the leading kinetic contribution to
$\Gammak^{(2)}$ and $R_k$ in eq. (\ref{twosix}), into an effective
inverse propagator (for massless fields)
\be
P(q^2) = Z_k q^2 + R_k = \frac{Z_k q^2}{1 - f^2_k(q)}.
\label{twothirteen} \ee
For $q^2 \gg k^2$ the inverse ``average'' propagator $P(q)$ approaches
the standard inverse propagator $Z_k q^2$ exponentially fast, whereas
for $q^2 \ll k^2$ the infrared cutoff guarantees $P(0) = Z_k k^2$.
Throughout most of this paper, we shall
work with the approximation of
$q^2$ independent $Z_k(\rho)$ and use the definition of eq. (\ref{twotwelve}).
When the $q^2$ dependence of
$Z_k(\rho,q^2)$ is considered, eq. (\ref{twotwelve}) has to be replaced by an
appropriate generalization, for example
\be
Z_k = Z_k(\rhzk,q^2=0).
\label{twofourteen} \ee
\par
Having summarized the basic formalism, we obtain in the next section
the exact evolution equation for the effective average potential
$U_k$.

\setcounter{equation}{0}
\renewcommand{\theequation}{{\bf 3.}\arabic{equation}}

\section*{3.
Exact evolution equation for the effective average
potential and its scaling form}

The evolution equation for $U_k(\rho)$ can be obtained by
calculating the trace in eq. (\ref{twosix}) for small field fluctuations
around a constant background configuration
$\phi_a(q=0) = \phi \delta_{a1},~~ \rho = \frac{1}{2} \phi^2$.
We define
\be
\phi_a(q) = \phi \delta_{a1} + \chi_a(q)
\label{threeone} \ee
and find \cite{exact}
\be
(\Gammak^{(2)})_{ab}(q,q') = \left[
\left( Z_k(\rho,q^2) \delta_{ab} +
\rho Y_k(\rho,q^2) \delta_{a1} \delta_{b1} \right) q^2
+ M^2_{ab}
\right] \delta(q-q'),
\label{threetwo} \ee
where the mass matrix $M^2_{ab}$ reads
\be
M^2_{ab} = U'_k(\rho) \delta_{ab} + 2 \rho U''_k(\rho) \delta_{a1} \delta_{b1}.
\label{threethree} \ee
Here primes denote derivatives with respect to $\rho$.
Eq. (\ref{twosix}) now reads
\be
\frac{\partial}{\partial t} U_k(\rho) =
\frac{1}{2} \int \frac{d^d q}{(2 \pi)^d}
\frac{\partial}{\partial t} R_k(q)
\left( \frac{N-1}{M_0} +\frac{1}{M_1} \right),
\label{threefour} \ee
with $R_k(q)$ given by eqs. (\ref{twotwo}) - (\ref{twothree}),
\beq
M_0(\rho,q^2) =~&Z_k(\rho,q^2) q^2 + R_k(q) + U'_k(\rho) \nonumber \\
M_1(\rho,q^2) =~&{\tilde Z}_k(\rho,q^2)  q^2
+ R_k(q) + U'_k(\rho) + 2 \rho U''_k(\rho)
\label{threefive} \eeq
and
\be
{\tilde Z}_k(\rho,q^2)  = Z_k(\rho,q^2) + \rho Y_k(\rho,q^2).
\label{threesix} \ee
For $\rho$ different from zero it is easy to recognize the
first term in eq. (\ref{threefour}) as the contribution from the
$N-1$ Goldstone bosons ($U'$ vanishes at the minimum).
The second term is then related to the radial mode.
The effective average potential $U_k$ has a $k$ dependent minimum
$\rhzk$.
We introduce the wave function renormalizations for the Goldstone and
radial modes (in agreement with eq. (\ref{twofourteen}))
\beq
Z_k =~&Z_k(\rhzk,q^2=0) \nonumber \\
Y_k =~&Y_k(\rhzk,q^2=0) \nonumber \\
\Zt_k =~&\Zt_k(\rhzk,q^2=0) = Z_k + \rhz(k) Y_k
\label{threesevenex} \eeq
When the $q^2$ dependence of
$Z_k(\rho,q^2)$,
$\Zt_k(\rho,q^2)$,
$Y_k(\rho,q^2)$
is neglected the wave function renormalizations are defined according to
eq. (\ref{twotwelve}) and read
\beq
Z_k =~&Z_k(\rhzk) \nonumber \\
Y_k =~&Y_k(\rhzk) \nonumber \\
\Zt_k =~&\Zt_k(\rhzk) = Z_k + \rhz(k) Y_k.
\label{threeseven} \eeq
The $k$ dependence of these functions is given by the anomalous
dimensions
\beq
\eta(k) =~&-\frac{d}{dt} (\ln Z_k) \nonumber \\
\etat(k) =~&-\frac{d}{dt} (\ln \Zt_k).
\label{threeeight} \eeq
At this stage the r.h.s. of eq. (\ref{threefour}) depends explicitely
on the scale $k$ once the momentum integration is performed.
The aim of this paper is a study of the critical behaviour at and near the
second order phase transition. At the phase transition one expects
a scaling behaviour of the effective average action $\Gamma_k$ and,
in particular, of the effective average potential $U_k$.
This holds provided approximate renormalized dimensionless couplings are
used. By a proper choice of variables the evolution equation (\ref{threefour})
should therefore be cast into a form where the scale $k$ no longer appears
explicitely.
In this formulation it will become easy to study the fixed point behaviour.
We perform the transformation of eq. (\ref{threefour}) into its scaling
form in several steps. First we introduce
the variable
$x=q^2$ and define the dimensionless functions
\beq
r_k(x) =~&\frac{R_k(x)}{Z_k x} \nonumber \\
s_k(x) =~&\frac{1}{Z_k} \frac{\partial}{\partial t}
\left(\frac{R_k(x)}{x} \right) =
\frac{\partial}{\partial t} r_k(x) +
\frac{\partial}{\partial t}(\ln Z_k) r_k(x) =
-2 x \frac{\partial}{\partial x} r_k(x)
-\eta r_k(x).
\label{threenine} \eeq
Here $R_k(x)/Z_k x$ is a dimensionless function of the ratio
$x/k^2$ (a property of $R_k$ independent of the specific choice
(\ref{twotwo})).
Eq. (\ref{threefour}) can then be written in the form
\be
\frac{\partial}{\partial t} U_k(\rho)
= v_d \int_0^{\infty} dx x^{\frac{d}{2}}
s_k
\left( \frac{N-1}{M_0/Z_k} +\frac{1}{M_1/Z_k} \right),
\label{threeten} \ee
with
\be
v_d^{-1} = 2^{d+1} \pi^{\frac{d}{2}} \Gamma \left( \frac{d}{2} \right).
\label{threeeleven} \ee
The ratios $M_0/Z_k$ and $M_1/Z_k$ read
\beq
M_0(\rho,x)/Z_k =~&z_k(\rho,x) x + r_k(x) x + Z_k^{-1} U'_k(\rho) \nonumber \\
M_1(\rho,x)/Z_k =~&\zh \zt_k(\rho,x) x + r_k(x) x + Z_k^{-1} U'_k(\rho)
+2 Z_k^{-1} \rho U''_k(\rho),
\label{threetwelve} \eeq
where
\beq
z_k(\rho,x) =~&Z_k(\rho,x)/Z_k \nonumber \\
\zt_k(\rho,x) =~&\Zt_k(\rho,x)/\Zt_k
\label{threethirteen} \eeq
and
\be
\zh(k) = \Zt_k/Z_k.
\label{threefourteen} \ee
This suggests the use of a dimensionless renormalized variable
\be
\rht=Z_k k^{2-d} \rho.
\label{threefifteen} \ee
In terms of $\rht$ one obtains
\be
\left. \frac{\partial U_k(\rht)}{\partial t} \right|_{\rht} =
(d-2+\eta)\rht \frac{\partial U_k(\rht)}{\partial \rht} +
v_d k^{-2} \int_0^{\infty} dx x^{\frac{d}{2}}
s_k
\left( \frac{N-1}{m_0} +\frac{1}{m_1} \right),
\label{threesixteen} \ee
where we have used the identity
\beq
\left. \frac{\partial U_k}{\partial t} \right|_{\rht}
=~&\left. \frac{\partial U_k}{\partial t} \right|_{\rho}
+ \left. \frac{\partial U_k}{\partial \rho} \right|_{t}
\left. \frac{\partial \rho}{\partial t} \right|_{\rht}
=
\left. \frac{\partial U_k}{\partial t} \right|_{\rho} +
\rht \left. \frac{\partial U_k}{\partial \rht} \right|_{t}
\left. \frac{\partial \ln \rht}{\partial \ln \rho} \right|_{t}
\left. \frac{\partial \ln \rho}{\partial t} \right|_{\rht}
\nonumber \\
=~&
\left. \frac{\partial U_k}{\partial t} \right|_{\rho}
+(d - 2 + \eta)\rht
\left. \frac{\partial U_k}{\partial \rht} \right|_{t}.
\label{threeseventeen} \eeq
The dimensionless ratios $m_0$ and $m_1$ are given by
\beq
m_0(\rht,x) =~&\frac{M_0}{Z_k k^2} = \frac{x}{k^2}
\left[ z_k(\rht,x)+r_k(x) \right] + k^{-d}
\frac{\partial U_k(\rht)}{\partial \rht}
\nonumber \\
m_1(\rht,x) =~&\frac{M_1}{Z_k k^2} = \frac{x}{k^2}
\left[ \zh \zt_k(\rht,x)+r_k(x) \right]
+ k^{-d} \left[
\frac{\partial U_k(\rht)}{\partial \rht}
+2 \rht
\frac{\partial^2 U_k(\rht)}{\partial \rht^2}
\right].
\label{threeeighteen} \eeq
We finally switch to a dimensionless potential
\be
u_k(\rht) = k^{-d} U_k(\rht)
\label{threenineteen} \ee
and use a dimensionless momentum variable
\be
y = x/k^2.
\label{threetwenty} \ee
We adopt the convention that the prime on $u_k(\rht)$
denotes a derivative with respect to $\rht$ at fixed $t$, while
the prime on $U_k(\rho)$ denotes a derivative with
respect to $\rho$ at fixed $t$.
At this point it is useful to fix the notation which we shall use in
the next sections. We denote the minimum of $U_k(\rho)$ with
$\rhz(k)$ and the minimum of $u_k(\rht)$ with $\kappa(k) = \rht_0(k)$.
For $\rhz(k) \not= 0$ (spontaneously broken regime) we have
$U'_k(\rhz)=0$, $u'_k(\kx)=0$. We define
$U''_k(\rhz)=\lb(k)$, $U^{(n)}_k(\rhz)=U_{n}(k)$ and
$u''_k(\kx)=\lx(k)$, $u^{(n)}_k(\kx)=u_{n}(k)$.
For $\rhz(k) = 0$ (symmetric regime) we define
$U'_k(0)=\mb(k)$,
$u'_k(0)=\mx(k)/k^2 = \mt(k)$ and as above for the higher derivatives.
We point out the relations
\beq
\kx =~&Z_k k^{2-d} \rhz
\nonumber \\
\mt =~&Z_k^{-1} k^{-2} \mb = k^{-2} \mx
\nonumber \\
\lx =~&Z_k^{-2} k^{d-4} \lb
\nonumber \\
u_n =~&Z_k^{-n} k^{(n-1)d - 2n} U_n.
\label{eextrae} \eeq
\par
We can now write the evolution equation in a scale independent form
\be
\frac{\partial}{\partial t} u_k(\rht) =
-d u_k(\rht) +(d-2+ \eta)\rht u'_k(\rht) +
v_d \int_0^{\infty} dy y^{\frac{d}{2}}
s(y)
\left( \frac{N-1}{m_0} +\frac{1}{m_1} \right),
\label{threetwentyone} \ee
with
\beq
m_0(\rht,y) =~&y \left[ z_k(\rht,y) + r(y) \right] + u'_k(\rht) \nonumber \\
m_1(\rht,y) =~&y \left[ \zh \zt_k(\rht,y) + r(y) \right] + u'_k(\rht)
+ 2 \rht u''_k(\rht).
\label{threetwentytwo} \eeq
Indeed, $r_k=r(y)$ only depends on $y$, whereas
$s_k=s(y, \eta) = -2 y \frac{\partial}{\partial y} r(y) - \eta r(y)$
depends on $y$ and $\eta$. We have, therefore, dropped the label $k$ for
these functions.
\par
Eq. (\ref{threetwentyone}) is the scaling form of eq. (\ref{threefour})
we were looking for. The behaviour exactly at the second order phase
transition should be given by a $k$-independent solution of this
equation.
The scaling solution of (\ref{threetwentyone}) obtains therefore for
${\partial u_k}/{\partial t}  = 0$.
Since no explicit $k$ dependence is present on the
r.h.s.
of (\ref{threetwentyone})
we conclude that
all the $k$ dependent quantities in
(\ref{threetwentyone}) obtain constant (fixed point) values.
In particular $\eta(k)$, $\etat(k)$ and
$\zh(k)$ take values $\eta_*$, $\etat_*$ and
$\zh_*$
for this solution. The definitions (\ref{threeeight}),
(\ref{threefourteen}) then imply
$\eta_*=\etat_*$.
Similarly $u_*(\rht)$ and $z_*(\rht,y)$, $\zt_*(\rht,y)$
become functions which are independent of $k$.
The nature of the fixed point becomes apparent
through the observation that the minimum of the potential appears for a
constant value
\be
\kappa(k) = \kappa_*.
\label{threetwentythree} \ee
As a result, for the fixed point solution, the minimum of the
effective potential is given by
\be
\rhz = \lim_{k\rightarrow 0} \rhz(k) \sim \lim_{k \rightarrow 0}
k^{d-2+\eta_*} \kappa_* = 0.
\label{threetwentyfour} \ee
We conclude that the fixed point of the evolution equation
indeed corresponds to the
phase transition between the spontaneously broken and the symmetric regime.
The scaling form of the potential $u_*(\rht)$ is determined by the equation
\be
d u_*(\rht) -( d-2+ \eta_*)\rht u'_*(\rht) +
2 v_d \int_0^{\infty} dy y^{\frac{d}{2}}
\left( y \frac{\partial r(y)}{\partial y} + \frac{\eta_*}{2} r(y) \right)
\left( \frac{N-1}{m_{0*}} +\frac{1}{m_{1*}} \right) = 0,
\label{threetwentyfive} \ee
with
\beq
m_{0*}(\rht,y) =~&y \left[ z_*(\rht,y) + r(y) \right] + u'_*(\rht) \nonumber \\
m_{1*}(\rht,y) =~&y \left[ \zh_* \zt_*(\rht,y) + r(y) \right] + u'_*(\rht)
+ 2 \rht u''_*(\rht).
\label{threetwentysix} \eeq
If the functions $z_*(\rht,y)$, $\zt_*(\rht,y)$ and the constants
$\eta_*$, $\zh_*$ are known
and the $y$-integration is performed, eqs. (\ref{threetwentyfive}),
(\ref{threetwentysix}) are reduced to a
non-linear second order differential equation for
$u_*(\rht)$.
\par
We point out that eqs. (\ref{threetwentyone}), (\ref{threetwentyfive})
are exact, since eq. (\ref{twoeleven}) contains the most general
terms which contribute to
$\Gamma^{(2)}_k$ when evaluated at a constant background field.
We have, therefore, obtained an exact non-perturbative
evolution equation which does not involve any explicit mass scale.

\setcounter{equation}{0}
\renewcommand{\theequation}{{\bf 4.}\arabic{equation}}

\section*{4.
Approximation with uniform wave function renormalization}

The solution $u_*(\rht)$ of the scaling equation
(\ref{threetwentyfive}) together with the evolution equation
(\ref{threetwentyone}) contains all the important information about
the behaviour at and near the phase transition. Indeed,
eq. (\ref{threetwentyone}) gives the functional dependence of the
various $\beta$-functions on the couplings at the phase transition.
(In fact, $u_*(\rht)$ describes infinitely many couplings and
eq. (\ref{threetwentyone}) specifies infinitely many
$\beta$-functions.) This allows to compute derivatives of the
$\beta$-functions with respect to the couplings. From there one
may compute the matrix of anomalous dimensions and infer the
critical exponents at the phase transition.
(For example, the critical exponent $\nu$ is related to the anomalous
mass dimension.) Even though eqs. (\ref{threetwentyone}) and
(\ref{threetwentyfive}) are exact the r.h.s. of
eq. (\ref{threetwentyfive}) still involves the unknown constants
$\eta_*$, $\zh_*$
and functions $z_*(\rht,y)$, $\zt_*(\rht,y)$.
If these are given one may attempt to compute the $y$-integral
in eq. (\ref{threetwentyfive}) numerically.
The result is a function depending on
$\rht$, $u'_*(\rht)$ and $u''_*(\rht)$.
The remaining non-linear second order differential equation can also be
treated with numerical methods. This program requires some effort, but
the result - exact values of the critical exponents - would be rewarding.
\par
Unfortunately, exact information on $\eta_*$, $\zh_*$, $z_*(\rht,y)$
and $\zt_*(\rht,y)$ is not available and the problem seems only
postponed. However, the situation is not so hopeless,
at least for the three-dimensional models
for which
$\eta_*$ is typically a small quantity
and $z_*(\rht,y)$, $\zt_*(\rht,y)$
are expected to depend only weakly on $\rht$ and $y$.
Then, a relatively crude approximation for
$\eta_*$, $\zh_*$, $z_*(\rht,y)$, $\zt_*(\rht,y)$,
which may be obtained by
truncating the exact evolution equation for these quantities or by
alternative methods, should already give excellent results. We shall
demonstrate the power of our method by making the very
simple approximation of uniform wave function renormalization, i.e.
\be
z_k(\rht,y)=\zt_k(\rht,y) = 1.
\label{fourone} \ee
This corresponds to neglecting the field and
momentum dependence of the wave
function renormalization. We observe that the dominant scale
dependence of the wave function renormalization - the anomalous dimension -
is always included by the $k$-dependence of $Z_k$ and $\Zt_k$
(the definition (\ref{threeseven}) applies).
For non-zero values of the infrared cutoff $k$ the more general
wave function renormalization function $Z_k(\rho,q^2)$ should be analytic
at $q^2 = 0$ even at the phase transition. This is a consequence of the
presence of the infrared cutoff. One expects a very mild momentum
dependence of $Z_k(\rhz,q^2)$ for $q^2$ between zero and $k^2$ whereas for
$q^2 \gg k^2$ the anomalous dimension should determine the behaviour near
the phase transition ($Z_k(\rhz,q^2) \sim (q^2)^{-\frac{\eta}{2}}$).
We note, however, that all momentum integrals are dominated by
$q^2 \simeq k^2$. Similarly, an expansion of $Z_k(\rho,q^2)$
in the field $\rho$ should be analytic at
$\rhz(k)$ as long as $k$ is strictly positive. (See section 6 for more
details.) Our approximation (\ref{fourone}) should, therefore, correctly
describe all qualitative features at the phase transition and be a good
quantitative approximation. Of course, if a high precision calculation
is attempted one will have to go beyond the truncation
(\ref{fourone}).
In the approximation (\ref{fourone})
it is useful to define the integrals
\beq
\Lt^d_0(\wt,\eta,\zh) =
{}~&2 \int^{\infty}_0 dy y^{\frac{d}{2}+1}
\frac{\partial r(y)}{\partial y}
\left[ y(\zh+r(y)) + \wt \right]^{-1}
\nonumber \\
&+ \eta \int^{\infty}_0 dy y^{\frac{d}{2}}
r(y)
\left[ y(\zh+r(y)) + \wt \right]^{-1}
\nonumber \\
\Lt^d_n(\wt,\eta,\zh) =
{}~&2n \int^{\infty}_0 dy y^{\frac{d}{2}+1}
\frac{\partial r(y)}{\partial y}
\left[ y(\zh+r(y)) + \wt \right]^{-(n+1)}
\nonumber \\
&+n \eta \int^{\infty}_0 dy y^{\frac{d}{2}}
r(y)
\left[ y(\zh+r(y)) + \wt \right]^{-(n+1)}~~~~~~~~~{\rm for}~~ n \geq 1,
\label{fourfour} \eeq
which obey
\beq
\frac{\partial}{\partial \wt} \Lt^d_0(\wt,\eta,\zh) =~&-\Lt^d_1(\wt,\eta,\zh)
\nonumber \\
\frac{\partial}{\partial \wt} \Lt^d_n(\wt,\eta,\zh)
=~&-n \Lt^d_{n+1}(\wt,\eta,\zh)
{}~~~~~~~~~~~~~~~~~~~~~~~~~~~~~~~~~~~~~~
{\rm for}~~ n \geq 1.
\label{fourfive} \eeq
Similarly we introduce
\be
L^d_n(\wt,\eta) = \Lt^d_n(\wt,\eta,1).
\label{fourtwo} \ee
In the following, for notational simplicity, we shall not
write down explicitly the dependence of
$L^d_n$ and $\Lt^d_n$ on $\eta$ and $\zh$.
Also, for the practical calculations in section 9, we shall
omit the contribution to $L^d_n$, $\Lt^d_n$ proportional to $\eta$
(we shall study the critical three-dimensional theory, for which
$\eta_* \simeq 0 - 0.04$) and we shall set
$\zh=1$
(which is again a good approximation near the fixed point).
With these simplifications we need only the detailed form of
$L^d_n(\wt) = L^d_n(\wt,0)$. For the choice (\ref{twotwo}) - (\ref{twothree})
for $R_k(q)$ these integrals are discussed in appendix A.
We emphasize, however, that all our formulae remain valid for
$\eta \not= 0$ and $\zh \not= 1$, even though the dependence on these
parameters is suppressed in the notation.
\par
In terms of $L^d_n$, $\Lt^d_n$ eq. (\ref{threetwentyone}) reads
\be
\frac{\partial u}{\partial t} =
-d u +(d-2+ \eta)\rht u' -
v_d (N-1) L^d_0(u')
- v_d \Lt^d_0(u'+2 \rht u''),
\label{foursix} \ee
where, for simplicity, we drop the subscript $k$ of $u_k$ from now on.
The evolution equations for the derivatives of $u$ are easily obtained
by differentiating (\ref{foursix}) with respect to $\rht$:
\beq
\frac{\partial u'}{\partial t} =~&(-2 + \eta) u' +(d-2+ \eta)\rht u''
\nonumber \\
&+ v_d (N-1) u'' L^d_1(u')
+ v_d (3 u'' + 2 \rht u''') \Lt^d_1(u'+2 \rht u'')
\label{fourseven} \\
\frac{\partial u''}{\partial t} =~&(d -4 + 2 \eta) u'' +(d-2+ \eta)\rht u'''
\nonumber \\
&- v_d (N-1) \bigl\{ [u'']^2 L^d_2(u') - u''' L^d_1(u') \bigr\}
\nonumber \\
&- v_d \bigl\{ (3 u'' + 2 \rht u''')^2 \Lt^d_2(u'+2 \rht u'') -
(5 u''' + 2 \rht u^{(4)}) \Lt^d_1(u'+2 \rht u'') \bigr\}.
\label{foureight} \eeq
In the spontaneously broken regime,
at the minimum $\kappa(k)$
we have
$u'(\kappa)=0$, $\lx = u''(\kx)$ and $u_n=u^{(n)}(\kx)$.
The running of the minimum is obtained by
taking a total $t$-derivative of the condition $u'(\kx)=0$
\beq
\frac{d \kx}{d t}
&=~\beta_{\kx} = - [u''(\kx)]^{-1}
\left. \frac{\partial u'}{\partial t} \right|_{\rht=\kx}
\nonumber \\
&=~- (d-2+ \eta)\kx -v_d (N-1) L^d_1(0)
- v_d  \left( 3 + \frac{2 \kx u_3}{\lx} \right)
\Lt^d_1(2 \lx \kx).
\label{fournine} \eeq
Eq. (\ref{foursix}) remains a complicated partial differential equation
for two variables $t$ and $\rht$, even if $L^d_0$, $\Lt^d_0$ take a simple
functional form. If we consider $u$ in the vicinity of some fixed value of
$\rht$, e.g. $\rht = \kx$, we can parametrize $u$ by its derivatives at this
fixed value, e.g. $\lx$, $u_n$. The evolution equation
is then expressed as an infinite system of coupled differential equations
for the couplings $\lx,u_n$. In particular, the $\beta$-function for
$\lx$ reads
\beq
\frac{d \lx}{d t} =~&\beta_{\lx} =
\left. \frac{\partial u''}{\partial t} \right|_{\rht=\kx}
+u_3 \frac{d \kx}{d t}
\nonumber \\
=~&(d-4+2 \eta) \lx -v_d (N-1) \lx^2 L_2^d(0)
\nonumber \\
&-v_d (3 \lx + 2 \kx u_3 )^2 \Lt^d_2(2 \lx \kx)
+v_d \left(2 u_3 + 2 \kx u_4 - \frac{2 \kx u_3^2}{\lx} \right)
\Lt^d_1(2 \lx \kx)
\label{fourten} \eeq
and it involves the higher derivatives $u_3$ and $u_4$. In turn, the
$\beta$-function for $u_3$ contains $u_4$ and $u_5$ and so on
(see section 7 for the details).
It should be also pointed out that
the functions $L^d_n$, $\Lt^d_n$
automatically introduce a ``threshold'' behaviour
in the $\beta$-functions, which
leads to the vanishing of the contributions from the massive radial mode
at scales $k$
much smaller than its running mass.
\par
As a starting point for the
understanding of the solutions of (\ref{foursix}) we need first the
behaviour of the scaling solution for $\partial u/ \partial t = 0$.
The scaling equation
\be
-d u_* +(d-2+ \eta_*)\rht u'_* -
v_d (N-1) L^d_0(u'_*)
- v_d \Lt^d_0(u'_*+2 \rht u''_*) = 0
\label{foureleven} \ee
is a non-linear differential equation for $\rht$, which remains difficult to
solve analytically. Numerical work for its solution is in
progress \cite{progress}.
For the present paper we rather concentrate on analytical aspects and present,
in the next two sections, results for limiting cases.

\setcounter{equation}{0}
\renewcommand{\theequation}{{\bf 5.}\arabic{equation}}

\section*{5.
Solution of the evolution equation for the effective average potential
for $N \rightarrow \infty$}

The main results of this work will be obtained through use of
the evolution equation (\ref{foursix}) for the study of
the three-dimensional theory. This will require a certain
amount of numerical work. It is useful, therefore, to
have analytical results in some limiting case, in order to
obtain conceptual understanding and verification
of the numerical solution. This is the purpose of this section
in which we concentrate on the three-dimensional theory in the large $N$
limit \cite{largenn}.
For $N \rightarrow \infty$ the evolution equation
(\ref{foursix}) simplifies considerably.
The first simplification obtains from the dominance of the
Goldstone contributions in the evolution equation.
The second one results
from the vanishing of the anomalous dimension $\eta$.
(This is a known result \cite{zinn} which will be reproduced
in section 9). This means that we can set $Z_k =1$ and omit all terms
involving $\eta$ in our expressions.
It is more convenient to study the
equation (\ref{fourseven}) for $u'$. The potential can be obtained
at the end through simple integration.
In the large $N$ limit eq. (\ref{fourseven}) takes the form
($d=3$)
\be
\frac{\partial u'}{\partial t} =-2 u' + \rht u''
+ v_3 N u'' L^3_1(u').
\label{fiveone} \ee
For the choice of
$R_k$ of eqs. (\ref{twotwo}), (\ref{twothree})
the function $L^3_1(u')$ does not have a
simple analytical form.
We use the notation
of eq. (\ref{afour}) and and make, in this section, the approximation
(with $l^3_1$ a constant of order unity)
\beq
L^3_1(u') =~&-2 l^3_1 s^3_1(u') \nonumber \\
s^3_1(u') =~&(1+ u')^{-2}.
\label{fivetwo}  \eeq
Notice that the above choice preserves the correct behaviour of
$L^3_1(u')$ as discussed in appendix A: monotonicity,
existence of a pole at $u'=-1$,
and asymptotic dependence on $u'$ for $u' \rightarrow \infty$.
We emphasize that the choice of $R_k$ is arbitrary within some general
conditions \cite{exact}. It may be possible to find an $R_k$ such that
eqs. (\ref{fivetwo}) become exact, with a suitably chosen value of
$l^3_1$. (See also footnote on page 17.)
\par
Eq. (\ref{fiveone}) can now be written as
\be
\frac{\partial u'}{\partial t}
- \rht \frac{\partial u'}{\partial \rht}
+ \frac{N C}{(1+u')^2} \frac{\partial u'}{\partial \rht}
+2 u' = 0,
\label{fivethree} \ee
where $C= 2 v_3 l^3_1 = l^3_1 / 4 \pi^2$.
We have also kept explicitly the factor of $N$
for consistency checks through the $N$ dependence of the couplings.
The most general solution of the partial differential equation
(\ref{fivethree}) is given by the relations
\beq
&\frac{\rht}{\sqrt{u'}} - \frac{N C}{\sqrt{u'}}
-\frac{N C}{2} \frac{\sqrt{u'}}{1+u'}
+ \frac{3}{2} N C \arctan \left( \frac{1}{\sqrt{u'}} \right)
= F \left( u' e^{2t} \right)~~~~~~~~~{\rm for}~~u'>0
\label{fivefour} \\
&\frac{\rht}{\sqrt{-u'}} - \frac{N C}{\sqrt{-u'}}
+ \frac{N C}{2} \frac{\sqrt{-u'}}{1+u'}
- \frac{3}{4} N C
\ln \left( \frac{1-\sqrt{-u'}}{1+\sqrt{-u'}} \right)
= F \left( u' e^{2t} \right)~~~{\rm for}~~u'<0.
\label{fivefive} \eeq
The function $F$ is undetermined until boundary conditions are specified.
For $t=0$ ($k=\Lambda$), $U_k$ coincides with the classical
potential which we choose
\be
U_{\Lambda}(\rho) = V(\rho) = - {\bar \mu}^2_{\Lambda}
 \rho + \frac{{\bar\lx}_{\Lambda}}{2} \rho^2.
\label{fivesix} \ee
The boundary condition, therefore, reads
\be
u'(\rht,t=0) = -\mu^2_{\Lambda} + \lx_{\Lambda} \rht,
\label{fiveseven} \ee
where
\be
\mu^2_{\Lambda} = \frac{{\bar \mu}^2_{\Lx}}{\Lambda^2},~~~~~\lx_{\Lambda} =
\frac{{\bar \lx}_{\Lx}}{\Lambda}.
\label{fiveeight} \ee
This uniquely specifies $F$ and we obtain
\beq
\frac{\rht}{\sqrt{u'}} - &\frac{N C}{\sqrt{u'}}
-\frac{N C}{2} \frac{\sqrt{u'}}{1+u'}
+ \frac{3}{2} N C \arctan \left( \frac{1}{\sqrt{u'}} \right)
= \nonumber \\
&\frac{\sqrt{u'} e^t}{\lx_{\Lambda}}
+\left( \frac{\mu^2_{\Lx}}{\lx_{\Lx}} - N C \right) \frac{1}{\sqrt{u'} e^t}
-\frac{N C}{2} \frac{\sqrt{u'} e^t}{1+u'e^{2t}}
+ \frac{3}{2} N C \arctan \left( \frac{1}{\sqrt{u'}e^t} \right)
{}~~~~{\rm for}~~u'>0
\label{fivenine} \\
\frac{\rht}{\sqrt{-u'}} - &\frac{N C}{\sqrt{-u'}}
+ \frac{N C}{2} \frac{\sqrt{-u'}}{1+u'}
- \frac{3}{4} N C
\ln \left( \frac{1-\sqrt{-u'}}{1+\sqrt{-u'}} \right)
= \nonumber \\
&-\frac{\sqrt{-u'} e^t}{\lx_{\Lambda}}
+\left( \frac{\mu^2_{\Lx}}{\lx_{\Lx}} - N C \right) \frac{1}{\sqrt{-u'}e^t}
+\frac{N C}{2} \frac{\sqrt{-u'} e^t}{1+u' e^{2t}}
- \frac{3}{4} N C
\ln \left( \frac{1-\sqrt{-u'} e^t}{1+\sqrt{-u'} e^t} \right)
{}~{\rm for}~~u'<0.
\nonumber \\
{}~~&~~
\label{fiveten} \eeq
\par
Even though the approximation of eq. (\ref{fivetwo}) does not permit
quantitative accuracy in all respects,
eqs. (\ref{fivenine}), (\ref{fiveten}) contain all the
qualitative information for the non-trivial behaviour of
the three-dimensional theory.
There is a critical value for the minimum of the classical potential
\be
\kx(\Lambda) = \frac{\mu^2_{\Lambda}}{\lx_{\Lambda}} = \kx_{cr} = N C,
\label{fiveeleven} \ee
for which a scale invariant (fixed point)
solution is approached in the limit
$t \rightarrow -\infty$. This solution is given by the relations
\beq
&\frac{\rht}{\sqrt{u'_*}} - \frac{N C}{\sqrt{u'_*}}
-\frac{N C}{2} \frac{\sqrt{u'_*}}{1+u'_*}
+ \frac{3}{2} N C \arctan \left( \frac{1}{\sqrt{u'_*}} \right)
= \frac{3 \pi}{4} N C~~~~~~~~~{\rm for}~~u'_*>0
\label{fivetwelve} \\
&\frac{\rht}{\sqrt{-u'_*}} - \frac{N C}{\sqrt{-u'_*}}
+ \frac{N C}{2} \frac{\sqrt{-u'_*}}{1+u'_*}
- \frac{3}{4} N C
\ln \left( \frac{1-\sqrt{-u'_*}}{1+\sqrt{-u'_*}} \right)
= 0~~~~~~~~~{\rm for}~~u'_*<0.
\label{fivethirteen} \eeq
On the other hand, the general solution of the
scaling equation, which results from eq. (\ref{fivethree}) by setting
$\partial u'/\partial t = 0$,
is given by
\beq
&\frac{\rht}{\sqrt{u'_*}} - \frac{N C}{\sqrt{u'_*}}
-\frac{N C}{2} \frac{\sqrt{u'_*}}{1+u'_*}
+ \frac{3}{2} N C \arctan \left( \frac{1}{\sqrt{u'_*}} \right)
= c_+~~~~~~~~~~~{\rm for}~~u'_*>0
\label{fivetwelveee} \\
&\frac{\rht}{\sqrt{-u'_*}} - \frac{N C}{\sqrt{-u'_*}}
+ \frac{N C}{2} \frac{\sqrt{-u'_*}}{1+u'_*}
- \frac{3}{4} N C
\ln \left( \frac{1-\sqrt{-u'_*}}{1+\sqrt{-u'_*}} \right)
= c_-~~~~~{\rm for}~~u'_*<0,
\label{fivethirteenee} \eeq
with $c_+$, $c_-$ arbitrary constants.
It is remarkable that there is only one
solution of the scaling equation
which is
continuous and finite for all finite $\rht$.
This is exactly given by eqs. (\ref{fivetwelve}), (\ref{fivethirteen}).
This criterion specifies more generally
the selection of the two free integration constants
which are necessarily present for the general solution of
eq. (\ref{foureleven}).
\par
Eqs. (\ref{fivetwelve}), (\ref{fivethirteen})
describe a potential $u$ which has a minimum at a constant value
\be
\kx(k) = \kx_* = N C,
\label{fivefourteen} \ee
and, according to (\ref{threetwentyfour}), corresponds to the phase transition
between the spontaneously broken and the symmetric regime.
(The values for $\kx_{cr}$ and $\kx_{*}$ coincide, but this is accidental.)
For the second $\rht$-derivative of $u$ at the minimum
$\lx = u''(\kx)$ we find
\be
\lx(k) = \lx_* = \frac{1}{2 N C},
\label{fivefifteen} \ee
and similar fixed point values for the higher derivarives of $u$.
For $1 \ll \rht/N C \ll (3 \pi/4) e^{-t} $ the rescaled potential
$u$ has the form
\be
u'_*(\rht) = \left( \frac{4}{3 \pi N C} \right)^2 \rht^2.
\label{fivesixteen} \ee
Notice that the region of validity of eq. (\ref{fivesixteen}) extends to
infinite $\rht$ for $t \rightarrow - \infty$.
{}From eq. (\ref{fivesixteen}) with $t \rightarrow -\infty
(k \rightarrow 0)$ we obtain for the effective potential at the
phase transition
\be
U_*(\rho) = \frac{1}{3} \left( \frac{4}{3 \pi N C} \right)^2 \rho^3
\label{fivesixteenex} \ee
and a critical exponent
$\delta = 5$, in agreement with standard large $N$ results
\cite{zinn}. (For the definition of $\delta$ and detailed discussion
for non-zero $\eta$ see section 6.)
\par
Through eqs. (\ref{fivenine}), (\ref{fiveten})
we can also study solutions which deviate slightly from the
scale invariant one. For this purpose we define a classical potential
with a minimum
\be
\kx(\Lambda) = \kx_{cr} + N \dkl,
\label{fiveseventeen} \ee
with $|\dkl| \ll 1$.
We find for the minimum of the potential
\be
\kx(k) = \kx_* + N \dkl e^{-t},
\label{fiveeighteen} \ee
and for $\lx$
\be
\lx(k) = \frac{\lx_*}{1 + \left( \frac{\lx_*}{\lx_{\Lambda}}
-1 \right)
e^t }.
\label{fivenineteen} \ee
Eq. (\ref{fiveeighteen}) indicates that the minimum of the potential stays
close to the fixed point value $\kx_*$ given by
(\ref{fivefourteen}), for a very long ``time''
$|t| < -\ln |\dkl|$. For $|t| > -\ln |\dkl|$ it deviates from the fixed point,
either towards the phase with spontaneous symmetry breaking
(for $\dkl > 0$), or the symmetric one
(for $\dkl < 0$).
Eq. (\ref{fivenineteen}) implies an attractive fixed point for $\lx$,
with a value given by eq. (\ref{fivefifteen}).
Similarly the higher derivatives are attracted to their fixed point
values.
The full phase diagram corresponds to a second order phase transition.
For $\dkl > 0$ the system ends up in the phase with spontaneous
symmetry breaking, with an
order parameter given by
\be
\rhz = \lim_{k\rightarrow 0} \rhz(k) = \lim_{k \rightarrow 0}
k \kappa(k) = N \dkl \Lx.
\label{fivetwenty} \ee
This leads to a critical exponent
\be
\beta = \lim_{\dkl \rightarrow 0}
\frac{d \left[\ln \sqrt{\rhz} \right]}{ d \left[\ln \dkl \right]} = 0.5,
\label{fivetwentyone} \ee
again in agreement with standard large $N$ results
\cite{zinn}. (For a detailed discussion
of the exponents $\beta$, $\nu$, $\zeta$ see section 9.)
In this phase the renormalized
quartic coupling
approaches zero linearly with $k$
\be
\lx_R = \lim_{k \rightarrow 0} k \lx(k) =
\lim_{k \rightarrow 0}k \lx_* =0.
\label{fivetwentyoneex} \ee
The fluctuations of the Goldstone bosons
lead to an infrared free theory in the phase with
spontaneous
symmetry breaking.
For $\dkl < 0$, $\kx(k)$ becomes zero at a scale
\be
t_s = - \ln \left( \frac{\kx_*}{N |\dkl|} \right)
\label{fivetwentytwo} \ee
and the system ends up in the symmetric regime ($\rhz=0$).
{}From (\ref{fivenine}), in the limit
$t \rightarrow -\infty$, with $u',u'',u''' \rightarrow \infty$,
so that
$u'e^{2t} \sim |\dkl|^2$, $u''e^{t} \sim |\dkl|/N$, $u''' \sim 1/N^2$,
we find
\be
U(\rho) = U_0(\rho) = \left( \frac{4}{3 \pi C} \right)^2
\left[ [\dkl]^2 \Lx^2 \rho - \frac{1}{N} \dkl \Lx \rho^2 + \frac{1}{3 N^2}
\rho^3 \right],
\label{fivetwentythree} \ee
in qualitative agreement with previous studies \cite{zinn, largen}.
In particular, we emphasize that the critical behaviour of the
three-dimensional theory coincides with the behaviour of the
four-dimensional theory at non-vanishing temperature near the
critical temperature $T_{cr}$. Eq. (\ref{fivetwentythree})
yields for both the symmetric phase and the phase with spontaneous
symmetry breaking ($\rho \geq \rho_{min}$)
\be
U'(\rho_4) = \left( \frac{2}{3 l^3_1} \right)^2
\left( \frac{8 \pi \rho_4}{NT} + \frac{\pi}{3} \frac{T^2- \tcr^2}{T}
\right)^2,
\label{ouf} \ee
provided we
identify
\be
\dkl~\Lambda = \frac{1}{24} \frac{\tcr^2- T^2}{T}
\label{ouff} \ee
and use the four-dimensional normalization for the field
$\rho_4 = \rho T$. Up to a possible difference
\footnote{
Eq. (3.15) in ref. \cite{largen}
obtains for $l^3_1 = 2/3$. It differs from
eq. (\ref{ouf}) in the present paper by an overall factor if we insert
the value (\ref{athree}), since $l^3_1 = \sqrt{\pi} / 2$. This difference is
due to the approximation of eq. (\ref{fivetwo}).
Note that it may be possible to alter the definition of $R_k$ in
eqs. (\ref{twotwo}), (\ref{twothree}) so that eq. (\ref{fivetwo}) becomes
exact. Then $l^3_1$ should come out to be $2/3$ with this special choice.
}
in the overall normalization
of $U'$ this is exactly the high temperature result of
ref. \cite{largen}. One obtains
the following well known values for the critical exponents
$\nu$, $\zeta$ in the symmetric phase
\beq
\nu =~&\lim_{\dkl \rightarrow 0}
\frac{d \left[\ln m_R \right]}{ d \left[\ln |\dkl| \right]}
= \lim_{\dkl \rightarrow 0}
\frac{d \left[\ln \sqrt{U'(0)} \right]}{ d \left[\ln |\dkl| \right]} = 1
\label{fivetwentyfour} \\
\zeta =~&\lim_{\dkl \rightarrow 0}
\frac{d \left[\ln \lx_R \right]}{ d \left[\ln |\dkl| \right]}
= \lim_{\dkl \rightarrow 0}
\frac{d \left[\ln U''(0) \right]}{ d \left[\ln |\dkl| \right]} = 1.
\label{fivetwentyfive} \eeq
Finally we find
\be
\lim_{\dkl \rightarrow 0}
\frac{\lx_R}{m_R} = \frac{8}{3 \pi N C},
\label{fivetwentysix} \ee
a result connected to the resolution of the problem of infrared
divergences of perturbation theory \cite{parisi, transition, largen}.
\par
Another approach for the study of eq. (\ref{fivethree})
would make use of a parametrization
of $u$ in terms of its minimum $\kx$ and the derivatives at the minimum
$\lx$, $u_n$, as discussed at the end of the previous section.
For the approximation of eq. (\ref{fivetwo}),
and in the large $N$ limit, eqs. (\ref{fournine}),
(\ref{fourten}) read
\beq
\frac{d \kx}{d t} =~&- \kx + N C
\label{fivetwentyseven} \\
\frac{d \lx}{d t} =~&- \lx + 2 N C \lx^2.
\label{fivetwentyeight} \eeq
The solution of the above equations, for a classical potential
of the form of eq. (\ref{fivesix}), is given by eqs.
(\ref{fiveeighteen}), (\ref{fivenineteen}).
Notice, however, that the disappearance of the higher derivatives
of the potential in eqs.
(\ref{fivetwentyseven}), (\ref{fivetwentyeight})
is a very convenient simplification
appearing only for $N \rightarrow \infty$.
In the next sections, where we shall study the full eqs.
(\ref{fournine}),
(\ref{fourten}) for small values of $N$,
more elaborate approximation schemes will be necessary.
\par
As a final comment we point out that, for a theory with spontaneous
symmetry breaking , we can use eq. (\ref{fiveten})
in order to study the ``inner'' part of the potential.
In particular, for $\rht = 0$ and $t \rightarrow -\infty$
eq. (\ref{fiveten}) predicts a potential $u$ which asymptotically behaves
as
\be
\lim_{t \rightarrow -\infty} u'(0) = -1.
\label{fivetwentynine} \ee
This leads to an effective average potential $U_k$ which becomes convex with
\be
\lim_{k \rightarrow 0} U'_k(0) = - k^2,
\label{fivethirty} \ee
in agreement with the detailed study of ref. \cite{convex}.

\setcounter{equation}{0}
\renewcommand{\theequation}{{\bf 6.}\arabic{equation}}

\section*{6.
Scaling solution for $k \rightarrow 0$ at fixed $\rho$}

We now turn back to the finite values of $N$.
The scaling equation (\ref{foureleven})
takes a particularly simple form for $\rht \rightarrow \infty$ which we
will study in this section. We observe that
the integrals $L^d_0(\wt)$, $\Lt^d_0(\wt)$ rapidly decrease for
$\wt \gg 1$ (see appendix A). For large values of $\rht$ we, therefore,
obtain the asymptotic equation
\be
-d u_* + (d-2 + \eta_*) \rht u'_* = 0.
\label{sixone} \ee
This has the simple general solution
\be
u_*(\rht) = c \rht^{\tau}
\label{sixtwo} \ee
with
\be
\tau = \frac{d}{d-2+ \eta_*}
\label{sixthree} \ee
and $c$ an integration constant.
\par
The limit $k \rightarrow 0$ for fixed $\rho$ corresponds to
$\rht \rightarrow \infty$ for $d-2 + \eta_* > 0$ (see eq.
(\ref{threefifteen})).
In this limit $U_k$ becomes the effective potential $U_0=U$.
As a result eq. (\ref{sixtwo}) gives the scaling form of
$U(\rho)$, namely
\be
U_*(\rho) = c Z_c^{\tau} \rho^{\tau}.
\label{sixfour} \ee
Here the constant $Z_c$ appears in the scaling form of the wave
function renormalization
\be
Z_k = Z_c k^{- \eta_*}.
\label{sixfive} \ee
The form of the potential (\ref{sixfour}) is directly related to
the critical exponent $\delta$ characterizing the relation between the
magnetization $\phi$ and the magnetic field
$B = \frac{\partial U}{\partial \phi} = \phi
\frac{\partial U}{\partial \rho}$
\be
B \sim \phi^{2 \tau -1} \sim \phi^{\delta}.
\label{sixsix} \ee
One obtains the well known scaling law
\be
\delta = \frac{d +2 -\eta_*}{d -2 +\eta_*}.
\label{sixseven} \ee
\par
At this point the reader may be somewhat surprised that the
simple approximation
(\ref{sixone}) automatically produces the correct relation
between $\delta$ and $\eta_*$. One would
expect that for $k^{d-2}$ much smaller than
$Z_k \rho$
the wave function renormalization $Z_k(\rho)$ becomes independent of $k$,
since no low mass modes are present. The identification of
$Z_k(\rho)$ with $Z_k$ as defined in eq. (\ref{threeseven})
becomes then problematic.
Indeed, the presence of two different infrared cutoffs: $k^2$ and
\be
m^2 = Z_k^{-1}(\rho) \frac{\partial U}{ \partial \rho}
\label{sixeight} \ee
suggests a qualitative behaviour
\be
Z_k(\rho) \sim (k^2 + C m^2)^{-\frac{\eta_*}{2}}.
\label{sixnine} \ee
This implies that $Z_k(\rho)$ and $Z_k=Z_k(\rhz(k))$
differ substantially for $\rht \gg 1$. Nevertheless, the asymptotic form
of the scaling equation, given by eq. (\ref{sixone}), remains valid.
This can be seen by introducing a different scaling variable
\be
\rhb = Z_k(\rho) k^{2-d} \rho
\label{sixten} \ee
which absorbs the dominant effects of the wave function renormalization
for large $\rht$ as well.
Using the scaling behaviour given by eqs.
(\ref{sixfour}), (\ref{sixnine}) for large $\rhb$ we find
\beq
Z_k(\rho) \sim~&m^{-\eta_*} \sim \rho^{-\frac{\eta_*(\tau-1)}{2-\eta_*}}
\sim \rho^{-\frac{\eta_*}{d-2 + \eta_*}}
\label{sixeleven} \\
\rhb \sim~&k^{2-d} \rho^{\frac{d-2}{d-2 + \eta_*}}.
\label{sixtwelve} \eeq
We replace eq. (\ref{threeseventeen}) by
\beq
\left. \frac{\partial U_k}{\partial t} \right|_{\rhb}
=~&\left. \frac{\partial U_k}{\partial t} \right|_{\rho}
+ \left. \frac{\partial U_k}{\partial \rho} \right|_{t}
\left. \frac{\partial \rho}{\partial t} \right|_{\rhb}
=
\left. \frac{\partial U_k}{\partial t} \right|_{\rho} +
\rhb \left. \frac{\partial U_k}{\partial \rhb} \right|_{t}
\left. \frac{\partial \ln \rhb}{\partial \ln \rho} \right|_{t}
\left. \frac{\partial \ln \rho}{\partial t} \right|_{\rhb}
\nonumber \\
=~&
\left. \frac{\partial U_k}{\partial t} \right|_{\rho}
+(d - 2)\rhb
\left. \frac{\partial U_k}{\partial \rhb} \right|_{t}.
\label{sixthirteen} \eeq
The asymptotic solution obtains in a similar way as before
\be
u_*(\rhb) \sim \rhb^{\frac{d}{d-2}}
\label{sixfourteen} \ee
and coincides with eq. (\ref{sixfour}).
We conclude that, in a leading approximation, the scaling equation
(\ref{sixone}) remains valid for large $\rht$ despite the difference
between $Z_k(\rho)$ and $Z_k$.
\par
We finally observe that corrections to the asymptotic solution
can be obtained from eq. (\ref{foureleven}) by factoring out the
asymptotic behaviour
\be
\uh(\rht) = u(\rht) \rht^{-\tau}.
\label{sixfifteen} \ee
This leads to
\beq
(d-2 + \eta_*) \rht \uh'_* -
&v_d (N-1) \rht^{-\tau} L^d_0 \left(\uh'_* \rht^{\tau}+ \tau
\uh _* \rht^{\tau-1}
\right)
\nonumber \\
-&v_d \rht^{-\tau} \Lt^d_0 \left(2 \uh''_*\rht^{\tau+1}
+(1+4 \tau) \uh'_* \rht^{\tau}
+ \tau(2 \tau -1)\uh_* \rht^{\tau-1} \right) = 0.
\label{sixsixteen} \eeq
The integrals $L^d_0$, $\Lt^d_0$ can be expanded for large values of the
argument $\wt$. (For the asymptotic behaviour of $L^d_0$ for $\eta=0$ see
eq. (\ref{afive}) in appendix A.)
In this way one can compute the leading corrections to eq. (\ref{sixtwo})
for large $\rht$. We postpone this discussion for a future publication
\cite{progress}.

\setcounter{equation}{0}
\renewcommand{\theequation}{{\bf 7.}\arabic{equation}}

\section*{7. The $\eta$ expansion}

In three dimensions, where
\be
\tau = \frac{3}{1 + \eta_*}
\label{sevenone} \ee
we observe that, according to eq. (\ref{sixtwo}),
the third derivative of $u_*(\rht)$ goes to zero
for $\rht \rightarrow \infty$
since $\eta_*$ is positive
\be
u'''_*(\rht) = c \tau (\tau -1) (\tau-2) \rht^{-\frac{3 \eta_*}{1+\eta_*}}.
\label{seventwo} \ee
This also holds for the higher derivatives of $u_*$.
On the other hand we learn from eq. (\ref{sixfour}) that
$U'''_*(\rho \rightarrow 0)$ diverges at the phase transition.
This shows that the $\phi^6$ coupling plays an important role at and near the
phase transition.
Since the fourth and higher derivatives of $U_*$ also diverge as
$\rho \rightarrow 0$ one may suspect that all these higher derivatives have to
be included for a quantitative estimate of the properties near the
phase transition.
This would require a solution of eq. (\ref{foursix}) without the
possibility of truncating the corresponding system of
infinitely many differential equations for $u^{(n)}$
(see the discussion at the end of
section 4) by setting $u^{(n)}=0$
for $n$ higher than a certain integer.
\par
Fortunately, the situation is much better due to the smallness of the
anomalous dimension $\eta_*$. (For the
three-dimensional theory $\eta_* \simeq 0-0.04$, see section 9.) We note that
for $\eta_* = 0$ the fourth and higer derivatives of $u_*(\rht)$
vanish identically for $\rht \rightarrow \infty$.
Numerical studies of eq. (\ref{foureleven}) indicate that this is
approximately true for small values of $\rht$ as well \cite{progress}.
Moreover, infinitesimally close to the phase transition the system
spends arbitrarily long ``time'' $|t|=-\ln(k/\Lambda)$
near the fixed point, before
deviating towards the phase with
spontaneous symmetry breaking
or the symmetric
one.
As a result the value of $\eta$ which
is relevant for the dynamics is the fixed point value $\eta_*$.
There should, therefore, exist an expansion in the value
of $\eta$ -- the $\eta$ expansion -- where the derivatives
$u^{(n)}$ for $n \geq 4$ give only small corrections
$\sim \eta$. For $\rht \rightarrow \infty$ we observe the relations
\beq
\rht u^{(4)}(\rht) =~&- \frac{3 \eta}{1+ \eta} u'''(\rht) = - \ex u'''(\rht)
\label{seventhree} \\
\rht^2 u^{(5)}(\rht) =~&\ex(\ex+1) u'''(\rht)
\label{sevenfour} \eeq
and similar for higher derivatives. All derivatives
$u^{(n)}$ with $n \geq 4$
are of order $\ex \sim \eta$ and combinations of the type
$\rht^2 u^{(5)} + \rht u^{(4)}$ are even suppressed
by higher powers of $\ex$ according to
\be
\left( \rht \frac{\partial}{ \partial \rht} \right)^n u'''(\rht)
=(- \ex)^n u'''(\rht).
\label{sevenfive} \ee
A truncation of (\ref{foursix}) should give reliable results at least for small
values of $\eta_*$.
Let us now try to
exploit these properties
also near the minimum of $u$ and
device a systematic truncation of eq. (\ref{foursix}).
The evolution of $u',u''$ is given by equations (\ref{fourseven}),
(\ref{foureight}) respectively.
For the next two derivatives one finds
\beq
\frac{\partial u'''}{\partial t} =~&(2d - 6 + 3 \eta) u'''
+(d-2+ \eta)\rht u^{(4)}
\nonumber \\
&+v_d (N-1) \bigl\{ 2 [u'']^3 L^d_3(u') - 3 u'' u''' L^d_2(u')+
u^{(4)} L^d_1(u') \bigr\}
\nonumber \\
&+v_d \bigl\{ 2 (3u''+2 \rht u''')^3 \Lt^d_3(u'+2 \rht u'')
-3(3u''+2 \rht u''')(5u'''+2 \rht u^{(4)}) \Lt^d_2(u'+2 \rht u'')
\nonumber \\
&+(7 u^{(4)} + 2\rht u^{(5)}) \Lt^d_1(u'+2 \rht u'') \bigr\}
\label{sevensix} \\
\frac{\partial u^{(4)}}{\partial t} =~&(3d - 8 + 4 \eta) u^{(4)}
+(d-2+ \eta)\rht u^{(5)}
\nonumber \\
&-v_d(N-1) \bigl\{ 6 [u'']^4 L^d_4(u') - 12[u'']^2 u''' L^d_3(u')
+\left(4 u'' u^{(4)} + 3[u''']^2 \right) L^d_2(u') - u^{(5)} L^d_1(u')
\bigr\}
\nonumber \\
&-v_d \bigl\{ 6(3u''+2 \rht u''')^4 \Lt^d_4(u'+2 \rht u'')
-12(3u''+2 \rht u''')^2 (5 u'''+ 2 \rht u^{(4)})
\Lt^d_3(u'+2 \rht u'')
\nonumber \\
&+ \left[ 4(3 u''+2 \rht u''')(7u^{(4)} +2 \rht u^{(5)})
+3(5u'''+2 \rht u^{(4)})^2 \right] \Lt^d_2(u'+2 \rht u'')
\nonumber \\
&-(9 u^{(5)} + 2 \rht u^{(6)}) \Lt^d_1(u'+2 \rht u'') \bigr\}.
\label{sevenseven} \eeq
As discussed earlier we can use these equations for a Taylor
expansion of $u$ around some fixed value. We choose $\rho=
\rhz(k)$ if the minimum of $u$ is away from the origin and
$\rho=0$ otherwise.
\par
In the \underline{spontaneously broken regime} $(\rhz(k) \not= 0)$,
the evolution of the
minimum $\kx$ of $u(\rht)$ is given by eq. (\ref{fournine}).
We consider the couplings
$\lx = u''(\kx)$, $u_n= u^{(n)}(\kx)$ which obey the
evolution equations
\be
\frac{d u_n}{d t} = \beta_n =
\left. \frac{\partial u^{(n)}}{\partial t} \right|_{\rht=\kx}
+u_{n+1} \frac{d \kx}{d t}.
\label{seveneight} \ee
The running of $\lx$ is given by eq. (\ref{fourten}) and the next two couplings
evolve according to
\beq
\frac{d u_3}{d t} =~&\beta_3 = (2d-6+3 \eta )u_3 +
v_d (N-1) \bigl\{ 2 \lx^3 L^d_3(0) - 3 \lx u_3 L^d_2(0) \bigr\}
\nonumber \\
&+v_d \bigl\{ 2(3 \lx + 2 \kx u_3)^3 \Lt^d_3(2 \lx \kx)
-3 (3 \lx + 2 \kx u_3)(5 u_3 +2 \kx u_4) \Lt^d_2(2 \lx \kx)
\nonumber \\
&+ \left( 4 u_4 + 2 \kx u_5 -\frac{2 \kx u_3 u_4}{\lx} \right)
\Lt^d_1(2 \lx \kx) \bigr\}
\label{sevennine} \\
\frac{d u_4}{d t} =~&\beta_4 = (3d-8+4 \eta )u_4
-v_d (N-1) \bigl\{ 6 \lx^4 L^d_4(0)- 12\lx^2 u_3 L^d_3(0)
+(4 \lx u_4 + 3 u_3^2) L^d_2(0) \bigr\}
\nonumber \\
&-v_d \bigl\{ 6(3 \lx + 2 \kx u_3)^4 \Lt^d_4(2 \lx \kx)
-12( 3 \lx +2 \kx u_3)^2 (5 u_3 +2 \kx u_4) \Lt^d_3(2 \lx \kx)
\nonumber \\
&+ \left[ 4(3 \lx +2 \kx u_3)( 7 u_4 +2 \kx u_5) +
3 (5 u_3 +2 \kx u_4)^2 \right] \Lt^d_2(2 \lx \kx)
\nonumber \\
&- \left(6 u_5 + 2 \kx u_6 - \frac{2 \kx u_3 u_5}{\lx} \right)
\Lt^d_1(2 \lx \kx) \bigr\}.
\label{seventen} \eeq
We observe that eqs.
(\ref{fournine}), (\ref{fourten}),
(\ref{sevennine}), (\ref{seventen})
in leading order in $N$, i.e. keeping the contribution
$(n(d-2+\eta)-d) u_n$ and the term $\sim (N-1)$, can be
solved level by level. In fact,
$\beta_{\kx}$, $\beta_{\lx}$ involve only $\kx$ and $\lx$,
$\beta_3$ depends only on $\kx$, $\lx$, $u_3$ and
$\beta_4$ on $\kx$, $\lx$, $u_3$, $u_4$.
In this limit no truncation is necessary
for the determination of the infrared fixed point
$\kx_*$, $\lx_*$, $u_{3*}$, $u_{4*}$.
This behaviour generalizes to arbitrary $n$.
The problem arises from the last bracket in the expressions for
$\beta_3$, $\beta_4$ which involves the couplings $u_5$, $u_6$
which are undetermined at this stage.
The same problem shifts to higher couplings if higher $\beta_n$ are computed.
The function $\beta_n$ always involves the couplings $u_{n+1}$, $u_{n+2}$
in addition to $u_m$ with $m \leq n$.
At this place some truncation of the infinite system of coupled
differential equations is required.
We need some information on $u_5$ and $u_6$.
A first possibility is to put
all $u_n$ with $n > 4$ to zero (truncation I). A second choice
(truncation II) is suggested by eq. (\ref{sevenfive}).
We approximate
\beq
u_5=~&- \kx^{-1} u_4,~~~~u_6 = 2 \kx^{-2} u_4~~~~~~~~~~{\rm for}~~
0.01 < 2 \lx \kx < 100
\nonumber \\
u_5=~&u_6 = 0~~~~~~~~~~~~~~~~~~~~~~~~~~~~~~~~~~{\rm for}~~
2 \lx \kx < 0.01~~{\rm or}~~2 \lx \kx > 100.
\label{seveneleven} \eeq
As we shall see in section 9, at the fixed point
$ 2 \kx_* \lx_* \simeq 1$, independent of $N$. Through the
truncation II we approximate $u_5$, $u_6$ by expressions
inspired by the analysis of the scaling solution, for the part of the
evolution during which the system is near the fixed point. For the rest of the
evolution we neglect them.
Numerical values for the fixed point for the two different truncations are
given in section 9, where all the numerical results are presented.
\par
In the \underline{symmetric regime} $(\rhz(k) = 0)$
we use the couplings
$\mt=u'(0)$, $\lx=u''(0)$, $u_n=u^{(n)}(0)$.
Their evolution equations read (in the symmetric regime
$\zh=1$, see eq. (\ref{threefourteen}), and the $L^d_n$, $\Lt^d_n$
integrals coincide)
\beq
\frac{d \mt}{d t} =~&\beta_{m} = (-2 + \eta) \mt
+v_d (N+2) \lx L^d_1(\mt)
\label{seventwelve} \\
\frac{d \lx}{d t} =~&\beta_{\lx} = (d-4+2 \eta) \lx
-v_d (N+8) \lx^2 L^d_2(\mt) + v_d (N+4) u_3 L^d_1(\mt)
\label{seventhirteen} \\
\frac{d u_3}{d t} =~&\beta_{3} = (2d-6+3 \eta) u_3
+ 2 v_d (N+26) \lx^3 L^d_3(\mt)
- 3 v_d (N+14) \lx u_3 L^d_2(\mt)
\nonumber \\
&+ v_d (N+6) u_4 L^d_1(\mt)
\label{sevenfourteen} \\
\frac{d u_4}{d t} =~&\beta_{4} = (3d-8+4 \eta)u_4
- 6 v_d (N+80) \lx^4 L^d_4(\mt)
+ 12 v_d (N+44) \lx^2 u_3 L^d_3(\mt)
\nonumber \\
&- 4 v_d (N+20) \lx u_4 L^d_2(\mt)
- 3 v_d (N+24) u^2_3 L^d_2(\mt)
+ v_d (N+8) u_5 L^d_1(\mt).
\nonumber \\
{}~&~
\label{sevenfifteen} \eeq
For the numerical work we shall neglect $u_5$ in the above equations, an
approximation consistent with both truncations I and II.
\par
The last necessary ingredients for the numerical study of the evolution
equations are the anomalous dimensions $\eta$, $\etat$, which we compute
in the following section.

\setcounter{equation}{0}
\renewcommand{\theequation}{{\bf 8.}\arabic{equation}}

\section*{8.
Anomalous dimensions}

In this section we compute the anomalous dimensions $\eta$ and $\etat$
(defined in eqs. (\ref{threeeight})), as
well as the evolution equation for the ratio $\zh=\Zt_k/Z_k$
(defined in eq. (\ref{threefourteen})).
The starting point is again the exact evolution equation (\ref{twosix}),
with $\Gamma_k$ parametrized according to eq. (\ref{twoeleven}).
In order to obtain the evolution equation for $Z_k$ (for
$N \geq 2$), we consider a
background field configuration with a small momentum dependence given in
momentum space by
\be
\phi_1(0) = \phi,~~~~~~~~~\delta \phi_2(Q) = \delta \phi,
\label{eightone} \ee
with $\df \ll \phi$, and $\phi_a = 0$ for $a > 2$.
(We remind the reader that we are concentrating on real fields for
which $\phi(-Q) = \phi^*(Q)$.)
Inserting the above configuration into (\ref{twoeleven}) we obtain
\be
Z_k(\rho) =
Z_k(\rho, Q^2=0) =
\lim_{Q^2 \rightarrow 0} \frac{\partial}{\partial Q^2}
\frac{\delta \Gamma_k}{\delta(\df^* \df)}
(\phi; \df=0).
\label{eighttwo} \ee
Similarly the evolution equation for $\Zt_k$ is obtained from
eq. (\ref{twosix}) by considering a configuration
\be
\phi_1(0) = \phi,~~~~~~~~~\delta \phi_1(Q) = \delta \phi,
\label{eightthree} \ee
with $\df \ll \phi$, and $\phi_a = 0$ for $a \geq 2$.
Eq. (\ref{twoeleven}) then gives
\be
\Zt_k(\rho) =
\Zt_k(\rho, Q^2=0) =
\lim_{Q^2 \rightarrow 0} \frac{\partial}{\partial Q^2}
\frac{\delta \Gamma_k}{\delta(\df^* \df)}
(\phi; \df=0).
\label{eightfour} \ee
For the evaluation of the r.h.s. of eq. (\ref{twosix}) we need an
expansion of the effective average action around the configurations
(\ref{eightone}), (\ref{eightthree}). This calculation has been done in ref.
\cite{christof2}. The results are summarized in appendix B, where the
general expressions for $\eta$ and $\etat$ are also derived.
In this section we limit our discussion to the approximation of uniform
wave function renormalization introduced in section 4 (see eq.
(\ref{fourone})).
This approximation corresponds to
considering only the value of $Z_k(\rho)$, $\Zt_k(\rho)$
at the minimum $\rhz(k)$, while
neglecting their derivatives.
The relevant expressions are obtained by inserting the truncations
\beq
Z'_k(\rhz) =~&Z''_k(\rhz) = 0
\nonumber \\
\Zt'_k(\rhz) =~&Z'_k(\rhz) + Y_k(\rhz) + Y'_k(\rhz) \rhz = 0
\nonumber \\
\Zt''_k(\rhz) =~&Z''_k(\rhz) + 2 Y'_k(\rhz) + Y''_k(\rhz) \rhz = 0
\label{eightfive} \eeq
into eqs. (\ref{btwentytwo}) and (\ref{bthirtynine})
(where the primes denote $\rho$-derivatives).
It is convenient to introduce the dimensionless quantity
\be
\yz = Z_k^{-2} k^{d-2} Y_k
\label{eightsix} \ee
such that
\be
\zh = \frac{\Zt_k}{Z_k} = 1 + \kx \yz.
\label{eightseven} \ee
It should be noted that for $N=1$ there is only one kinetic invariant.
The wave function renormalization should
be identified with $\Zt_k$, while $Y_k$ can be set to zero.
\par
In the \underline{spontaneously broken regime}
one finds for the anomalous dimension $\eta$
\be
\eta = - \frac{2 v_d}{d} \kx^{-1} \Mh^d_2(2 \lx \kx)
-v_d \yz \left[ \Lt^d_1(2 \lx \kx) - \frac{2}{d} \yz \kx \Lt^{d+2}_2(2 \lx \kx)
\right]
\label{eighteight} \ee
and similarly for $\etat$
\beq
\etat =~&v_d (N-1) \yz \zh^{-1} \left[ L^d_1(0) + 2 \lx \kx L^d_2(0) \right]
\nonumber \\
&- \frac{4 v_d}{d} \zh^{-1} \kx
\left[ (N-1) \lx^2 Z_k^2 M^d_{4,0}(0) + (3 \lx + 2 u_3 \kx)^2 Z_k^2
\Mt^d_{4,0}(2 \lx \kx) \right].
\label{eightnine} \eeq
The functions $M^d_{4,0}$, $\Mt^d_{4,0}$ and $\Mh^d_2$
are defined in eqs. (\ref{bsixteen}), (\ref{bthirtyfive}) and
(\ref{btwentyseven}) respectively. Notice that in eqs.
(\ref{eighteight}), (\ref{eightnine}) we are using rescaled
arguments with the conventions of eq. (\ref{eextrae}).
We shall need the function $\Mh^d_2$
for the numerical calculations of the next section.
Within the approximation of uniform wave function renormalization
and with the notation introduced in section 3, it can be written as
\beq
\Mh^d_2(\wt,\eta,&\zh) = -2 \int_0^{\infty} dy y^{\frac{d}{2}}
\left( 1 + r +y \frac{\partial r}{ \partial y} \right)
\left[ \frac{1}{(1+r)y} - \frac{1}{(\zh+r)y +\wt} \right]^2
\nonumber \\
&\biggl\{
2 y \frac{\partial r}{ \partial y}
+ 2 \left(y \frac{\partial }{ \partial y} \right)^2 r
+ \eta r + \eta y \frac{\partial r}{ \partial y}
\nonumber \\
&- y \left( 1 +r + y \frac{\partial r}{ \partial y} \right)
\left( \eta r +2 y \frac{\partial r}{ \partial y} \right)
\left[ \frac{1}{(1+r)y} + \frac{1}{(\zh+r)y +\wt} \right]
\biggr\}.
\label{eighteleven} \eeq
(The dependence of $\Mh^d_2$
(and the other functions) on $\eta$ and $\zh$ has been suppressed
in the notation of eq. (\ref{eighteight}) for simplicity.)
This integral simplifies considerably if the contributions
proportional to $\eta$ are neglected and we set $\zh=1$.
With these simplifications it depends only on $\wt$ and is discussed
in the appendix C.
The evolution equation for $\zh$ can be expressed in terms of $\eta$ and
$\etat$
\be
\frac{d \zh}{dt} = \frac{d(\kx \yz)}{dt} = \zh (\eta - \etat).
\label{eightten} \ee
For the scaling solution $\zh$ is at its infrared stable fixed point
$\zh_*$ (see discussion at the end of section 3).
{}From eq. (\ref{eightten}) we conclude that the fixed point values
$\eta_*$ and $\etat_*$ coincide. In other words, $\zh$ adjusts
itself so that the two anomalous dimensions become identical.
This behaviour is expected to be a general feature, independent of
the particular truncation used in this paper.
\par
In the \underline{symmetric regime}, where $\rhz(k)=0$, we obtain from
eq. (\ref{bfourty})
\be
\eta = \etat = - v_d \yz L^d_1(\mt).
\label{eighttwelve} \ee
At the transition between the spontaneously broken and symmetric regime
(for $\kx=0$, $\mt=0$) this expression agrees with
eq. (\ref{eighteight}) but not with eq. (\ref{eightnine}).
The reason is that for small $\kx$ the truncation
$\Zt'_k(\rhz)=0$ ceases to be consistent with a nonvanishing
$Y_k(\rhz)$ (and $Z'_k(\rhz)=0$).
A reliable computation of $\etat$ for very small nonvanishing
$\kx$ should improve over the truncation of eq. (\ref{eightfive}).
For similar reasons the present truncation does not guarantee
$\zh \rightarrow 1$ for $\kx \rightarrow 0$ in the spontaneously
broken regime. Such a behaviour is required by
continuity in a treatment without truncations, since $\zh$ exactly
equals one in the symmetric regime. We note that beyond the
approximation $Z'_k(\rhz) = \Zt'_k(\rhz) = 0$ the values of $\eta$
or $\etat$ do not necessarily coincide at the transition between the
two regimes, since the definition of the anomalous dimensions
involves additional terms from the $k$ dependence of $\rhz(k)$ in the
spontaneously broken regime. For very small $\kx$ we may approximate
$\etat$ in the spontaneously broken regime by neglecting the
contribution from $\Zt^{(b)}$ of eq. (\ref{bthirtytwo}) and
setting
$Z'_k(\rhz) =0$, $Y'_k(\rhz) \rhz=0$, $U'''(\rhz) \rhz =0$
\be
\etat = v_d \zh^{-1} \yz
\bigl\{ 2 \Lt^d_1(2 \lx \kx) + (N-1) L^d_1(0)+ \frac{\yz}{\lx}
\Lt^{d+2}_1(2 \lx \kx) \bigr\}.
\label{eightthirteen} \ee
Finally we need the evolution equation for $\yz$ in the symmetric
regime
\beq
\frac{d \yz}{d t } =~& (d-2+2 \eta)  \yz + Z_k^{-2} k^{d-2} \frac{d Y_k}{ dt}
\label{eightfourteen} \\
\frac{d Y_k}{ dt} =~&\lim_{\rho \rightarrow 0} \frac{1}{\rho}
\frac{\partial }{ \partial t} \left[ \Zt_k(\rho) - Z_k(\rho) \right].
\label{eightfifteen} \eeq
Here ${\partial Z_k(\rho)}/{ \partial t}$ and
${\partial \Zt_k(\rho)}/{ \partial t}$
can be extracted from eqs. (\ref{bseven}), (\ref{beight}) and
(\ref{bthirtyone}), (\ref{bthirtytwo})
respectively, with $w_1=w_2 = U'_k(\rho)$ and $Z_k(\rho)=\Zt_k(\rho)$.
We observe that eq. (\ref{eightfifteen}) is well defined since
\be
\lim_{\rho \rightarrow 0}
\frac{\partial }{ \partial t}  \Zt_k(\rho) =
\lim_{\rho \rightarrow 0}
\frac{\partial }{ \partial t}  Z_k(\rho)  =
v_d k^{d-2} Z_k^{-1}
\left[ N Z'_k(0) + Y_k(0) \right]
L^d_1(\mt).
\label{eightsixteen} \ee
Due to the presence of massive modes,
$Y_k$ quickly stops running in the symmetric
regime and $\yz$ approaches zero for $d > 2$. Since in the present paper the
running in the symmetric regime plays only a secondary role, we shall use
in this regime a somewhat stronger truncation by setting
$Y_k(0) = 0$. This implies for the symmetric regime
\be
\eta = \etat =0.
\label{eightseventeen} \ee

\setcounter{equation}{0}
\renewcommand{\theequation}{{\bf 9.}\arabic{equation}}

\section*{9.
Critical exponents of the three-dimensional theory}

In this section we use the formalism developed in the previous sections
in order to study the phase transition of the
three-dimensional $N$-component $\phi^4$ theory.
As we have already discussed in section 2, the effective average action
interpolates between the classical action (for $k=\Lambda$)
and the effective action (for $k = 0$). Therefore, our strategy
is to define the classical theory at the ultraviolet cutoff scale and solve
the evolution equation for the effective average action (in some
appropriate truncation scheme),
in order to determine the renormalized theory in the limit
$k \rightarrow 0$.
Subsequently, we can adjust the classical (bare)
parameters of the theory in order to approach the fixed point
in the critical region and obtain
all the quantitative information for the phase transition.
\par
The evolution equations for the parameters of the effective average potential
are given by expressions (\ref{fournine}), (\ref{fourten}), (\ref{sevennine}),
(\ref{seventen}) for the spontaneously broken regime, and
(\ref{seventwelve}) - (\ref{sevenfifteen}) for the symmetric regime.
The derivatives of the potential higher than the fourth are either neglected
(truncation I and symmetric regime) or estimated according to
eq. (\ref{seveneleven}) (truncation II).
For the numerical calculation
we consider only the anomalous dimension for the Goldstone modes $\eta$.
We have shown in section 8 that, for the scaling (fixed point) solution, the
anomalous dimensions for the radial and Goldstone modes coincide:
$\etat_*=\eta_*$. Since in this work we are interested only in the critical
region, it is a good approximation to drop the distinction between the
two anomalous dimensions. For practical purposes we use eq.
(\ref{eighteight}), keeping only the first term and neglecting the ones
proportional to $y_0$. It should be noted that
this approximation is not valid for $N=1$, where the only anomalous
dimension is $\etat$. It is expected, however, from continuity of
our expressions in the parameter $N$, that
eq. (\ref{eighteight}) should give a good approximation to $\etat_*$
even for $N=1$.
In the symmetric phase we set $\eta=0$ according to
eq. (\ref{eightseventeen}).
For the numerical calculation
we also drop the distinction between the integrals $L^d_n$ and
$\Lt^d_n$ as explained in section 4 after eq. (\ref{fourtwo}).
In this approximation these integrals are discussed in appendix A.
Since they cannot be evaluated in closed form for a general argument, we
perform numerical fits of $L^3_n(\wt)$, which connect the expressions
(\ref{athree}) for $\wt=0$ with the asymptotic expressions
(\ref{afive}).
Another numerical fit is used for the integral $\Mh^3_2(\wt)$ which is
discussed in appendix C.
\par
A typical example of the numerical integration of the evolution
equations is displayed in fig. 2, for $N=3$. The truncation II is used,
even though the parameters $u_4$, $u_5$, $u_6$ are not plotted.
The evolution starts at $k=\Lambda$ with the classical action, so that
$Z_{\Lambda}=1$, $u_3(\Lambda)=u_4(\Lambda)=u_5(\Lambda)=u_6(\Lambda)=0$.
The quartic coupling is arbitrarily chosen
$\lx(\Lambda)=\lb(\Lambda)/\Lambda=0.1$. Two values of $\kx(\Lambda)$
are selected so that the system is very close to the two
sides of the critical line.
The two values are indistinguishable in the graph
and are given by
$\kx(\Lambda) = \kx_{cr} + \dkl$, with
$\kx_{cr} = [\rho_0(\Lambda)]_{cr}/\Lambda \simeq 0.107$,
$|\dkl| \ll \kx_{cr}$ and $\dkl$ positive or negative.
With these initial conditions the system evolves towards the fixed point in
the critical region and remains close to it for a long ``time''
$|t|=-\ln(k/\Lambda)$. The fixed point values of the different parameters
are listed in the last row of table 1. By staying near the fixed point for
several orders of magnitude in $t$, the system loses memory of the initial
conditions at the cutoff. As a result the critical behaviour is
indepenent of the detailed short distance physics and is
determined by the fixed point, as long as the evolution
starts sufficiently close to the critical line.
In the final part of the evolution, as $k \rightarrow 0$, the theory
settles down either in the phase with spontaneous symmetry breaking
or the symmetric one,
for $\dkl$ positive or negative respectively.\\
a) For positive $\dkl$, the physical
parameters of the theory approach their renormalized values
in the limit $k \rightarrow 0$.
Due to the presence of the mass for the radial
mode, the anomalous dimension
$\eta$ becomes zero for $k \rightarrow 0$, and the wave function
renormalization approaches a finite value
$Z_0= Z$.
We define
\beq
\rho_{0R} =~&\lim_{k \rightarrow 0} k \kx(k) = Z^{-1} \rhz
\nonumber \\
\lx_R =~&\lim_{k \rightarrow 0} k \lx(k) = Z^{-2} \lb_0
\nonumber \\
U_{nR} =~&\lim_{k \rightarrow 0} k^{3-n}  u_n(k)
= Z^{-n} U^{(n)}_0(\rhz).
\label{nineone} \eeq
For $N \not= 1$ and $k \rightarrow 0$, the quantity
$\lx(k)$ approaches a
finite value. As a result, due to the
fluctuations of the Goldstone modes,
the renormalized quartic coupling $\lx_R$ is zero in the
phase with spontaneous breaking and the theory is infrared free.
This is not the case for $N=1$, where $\lx$ diverges in the
infrared so that $\lx_R$ remains non-zero.
\\ b) For negative $\dkl$, the quantity
$\kx(k)$ becomes zero at some non-zero $k_s$.
The evolution is continued with the equations for the
symmetric regime, with
$\mt(k_s)=0$ and $u_n(k_s)$ assuming their values
at the end of the running in the spontaneously broken regime.
This part of the evolution is not displayed in fig. 2 for
simplicity. Notice that
the anomalous dimension approaches zero at $k_s$ and remains zero in the
symmetric regime.
For $k \rightarrow 0$ the physical parameters of the theory approach their
renormalized values
\beq
\mx_R =~&\lim_{k \rightarrow 0} k^2 \mt(k)
\nonumber \\
\lx_R =~&\lim_{k \rightarrow 0} k \lx(k)
\nonumber \\
U_{nR} =~&\lim_{k \rightarrow 0} k^{3-n}  u_n(k).
\label{ninetwo} \eeq
Very close to the phase transition,
$m_R$ and $\lx_R$ are very
small compared to $\Lambda$, with a fixed value for the ratio $m_R/\lx_R$.
We shall return to this point later in this section.
\par
In table 1 the fixed point values of various parameters are listed, for
$N=3$ and various
truncations of the evolution equations.
For the first row only two evolution equations are considered
for $\kx$ and $\lx$, and the higher derivatives of the potential
as well as the anomalous dimension are set to
zero. Then, the number of evolution equations is enlarged to three with the
inclusion of $u_3$. Subsequently $\eta$ and $u_4$ are added (truncation I), and
finally $u_5$ and $u_6$ are approximated according
to eq. (\ref{seveneleven}) and are included in the system of five
coupled evolution equations (truncation II).
It becomes apparent from table 1, that the influence of the higher
derivatives of the potential on the fixed point,
and therefore on the critical dynamics, is small.
In fact, for an efficient truncation it would suffice to consider the
``relevant'' first three $\rho$-derivatives of the potential and the anomalous
dimension. The inclusion of higher derivatives gives small improvements
in agreement with the analysis of section 7.
In table 2 we list the fixed point values of the various parameters
for several values of $N$. It is apparent that the anomalous dimension at the
fixed point is always a small quantity, and that
$2 \lx_* \kx_* \simeq 1$ for all $N$.
This justifies eq. (\ref{seveneleven}) for the truncation II.
\par
The critical exponents parametrize the singular behaviour
of the theory
as the phase transition is approached. A measure of the distance from the
phase transition is the small difference
\be
\dkl = \kx(\Lambda) - \kx_{cr}
\label{ninethree} \ee
for any given value of $\lx(\Lambda)$.
If $\kx(\Lambda)$ is interpreted as a function of temperature
in a realistic three-dimensional model, the deviation $\dkl$
is proportional to the deviation from the critical temperature,
$\dkl = A_{\kx} (\tcr- T)$.
The critical exponent $\beta$ parametrizes the behaviour of the
``unrenormalized'' order parameter
\be
\rhz = \lim_{k \rightarrow 0} \rhz(k)
= Z^{-1} \rho_{0R}
\label{ninefour} \ee
according to
\be
\rhz \sim (\dkl)^{2 \beta},
\label{ninefive} \ee
for $\dkl \rightarrow 0^+$.
Similarly, the critical exponent $\nu$ describes the divergence near the phase
transition of the correlation length, which is equal to the inverse of the
renormalized mass $m_R$.
As we have already mentioned, the theory is infrared free
in the phase with spontaneous breaking for $N \not= 1$.
This means that $\lx_R$ vanishes and, since $\rho_{0R}$ is fixed, the
renormalized mass $m_R$ vanishes
\footnote{
We could avoid this problem by defining $\lx_R$ and $m_R$ by renormalized
four and two point functions with non-vanishing external momenta.
}
As a result, for
$\dkl > 0$ the critical exponent $\nu$ can be defined only for $N=1$,
according to
\be
m_R = \sqrt{2 \lx_R~\rho_{0R}} \sim
(\dkl)^{\nu_-}.
\label{ninesix} \ee
For $\dkl < 0$ the theory is in the symmetric phase and, for any $N$, we can
define
\be
m_R \sim |\dkl|^{\nu_+}.
\label{nineseven} \ee
(Here $m_R$ is given by $m(k=0)$.)
The critical exponent $\gamma$ is related to the response of the expectation
value $\phi^a$ to an external magnetic field or source $B^a$.
The magnetic susceptibility
\be
\chi = \frac{\partial \phi^a}{\partial B^a},
\label{nineeight} \ee
evaluated for a vanishing magnetic field
$B^a = 0$,
is non-analytic
for $\dkl \rightarrow 0$
\be
\chi \sim |\dkl|^{-\gamma}.
\label{ninenine} \ee
The relation between $\phi^a$ and $B^a$
is encoded in the effective potential
and reads for small values of
$\phi^a$
and $\dkl \not= 0$
\beq
B^a =~&\frac{\partial U(\phi)}{\partial \phi_a} =  U' \phi^a =
\mb \phi^a \nonumber \\
\chi =~&\bar{m}^{-2}.
\label{nineten} \eeq
Here the ``unrenormalized'' mass term $\mb$ is given by
\be
\mb = \lim_{k \rightarrow 0} \mb(k) = Z \mx_R.
\label{nineeleven} \ee
For $N=1$ and $\dkl > 0$ we obtain
\be
\mb = 2 \lb_0 \rhz =
2 Z \lx_R~\rho_{0R} \sim
(\dkl)^{\gamma_-}.
\label{ninetwelve} \ee
In the symmetric phase
($\dkl < 0$) we have for arbitrary $N$
\be
\mb = Z m^2_R \sim |\dkl|^{\gamma_+}.
\label{ninethirteen} \ee
\par
For the numerical determination
of the critical exponents we integrate the
evolution equations with the truncation II, for several very small values
of $\dkl$. The critical exponents are then obtained as the slope of the
logarithm of the relevant quantity at $k=0$, when plotted as a function of
$\ln|\dkl|$. More specifically, for the determination of $\beta$ we
compute the slope of $\ln \rhz$ as a function of $\ln \dkl$, and similarly
for the other critical exponents.
The behaviour of the exponents for decreasing $|\dkl|$ is
depicted in
fig. 3 for $N=4$. As $\dkl \rightarrow 0$ they
approach universal values.
The deviation from the universal behaviour for
larger values of $\dkl$ depends on the details of the bare theory.
Fig. 3 corresponds to a bare theory (defined at $k = \Lambda$) with
$N=4$, $\lx(\Lambda)=0.5$, $\kx_{cr} \simeq 0.120$.
We have verified that the asymptotic values of the critical exponents
are universal by determining them for different
$\lx(\Lambda)$ and $\kx_{cr}$.
For the anomalous dimension the
relevant value is the fixed point value $\eta_*$.
The results of our calculation are summarized in table 3 for various
$N$. For $N=1$ the critical exponents $\nu$ and $\gamma$ satisfy
$\nu_+= \nu_-$ and $\gamma_+ = \gamma_-$ with an accuracy of 0.1\%. For
$N \not= 1$ we list the values for $\nu_+$ and $\gamma_+$.
The critical exponents satisfy the scaling laws
\beq
\beta =~&(1+ \eta_*) \frac{\nu}{2}
\label{ninefourteen} \\
\gamma =~&(2-\eta_*) \nu
\label{ninefifteen} \eeq
with an accuracy of 0.1\%.
We mention also the scaling law of eq. (\ref{sixseven}), which gives the
critical exponent $\delta$ and
was obtained in section 6.
In table 3 we have included values for the critical exponents obtained
through other methods (summed perturbation series in three dimensions,
$\ex$-expansion, lattice calculations and $1/N$-expansion),
for comparison.
We observe agreement at the 1-5 \% level for the exponents
$\beta$, $\nu$, and $\gamma$. For the anomalous dimension $\eta_*$
we observe satisfactory
agreement of our results with the quoted values, even though
$\eta_*$ is a small quantity and
most severely affected by the various approximations
in our method.
In fact, significant improvement for all the results is expected if the
anomalous dimension for the radial mode $\etat$ is
included. Preliminary results indicate that this requires
a truncation that includes at least the first $\rho$-derivative of the
wave function renormalizations. We postpone this discussion for a future
publication \cite{progress}.
An interesting point, which is
connected to the resolution of the problem of infrared divergences
of perturbation theory \cite{parisi, transition, largen},
concerns the vanishing of the renormalized quartic coupling $\lx_R$ at the
phase transition. More specifically, the ratio $\lx_R/m_R$ takes
an $N$ dependent value in the critical region. As a result, the
behaviour of $\lx_R$ in the critical region can be characerized by
a critical exponent
$\zeta$ defined according to
\be
\lx_R \sim |\dkl|^{\zeta}
\label{ninesixteen} \ee
and
\be
\zeta = \nu.
\label{ninetwenty} \ee
In table 4 we list our results
for the ratio $\lx_R/m_R$, as well as results obtained with other methods for
comparison. For $N \not= 1$, due to the vanishing of $\lx_R$ in the phase
with spontaneous symmetry breaking, the ratio was calculated in the symmetric
phase. For $N =1$, the values obtained in the phase with spontaneous
symmetry breaking and the symmetric one
agree at the 0.1 \% level.

\setcounter{equation}{0}
\renewcommand{\theequation}{{\bf 10.}\arabic{equation}}

\section*{10. Conclusions}
The effective average action $\Gamma_k$
results
from the effective integration of fluctuations with characteristic momenta
larger than a given infrared cutoff $k$.
It is the appropriate quantity for the study of the physics at the scale $k$.
It interpolates between the classical action $S$ for $k=\Lambda$ ($\Lambda$
being the ultraviolet cutoff of the theory, much larger than any other
physical scale) and the effective action $\Gamma$ for $k=0$.
The dependence of $\Gamma_k$ on $k$ is given by the exact evolution
equation (\ref{twosix}) \cite{exact}.
No ultraviolet or infrared divergences appear in this evolution equation,
and the formalism leads to an efficient treatment of the infrared problems
which plague theories with massless modes in less than four dimensions.
In this work we have
studied the exact non-perturbative evolution equation for the
effective potential. It is well suited for the study of critical phenomena
since it can be cast into a scale independent form. In particular,
a second order phase transition corresponds to fixed points for the
various couplings appearing in the potential. They can be computed
as the solution of a non-linear second order differential equation for the
field dependence of the potential. We have given the explicit solution
of this differential equation in the large $N$ limit
of the $O(N)$-symmetric scalar theory in three dimensions.
It describes the fixed points for infinitely many dimensionless couplings.
We also have solved the more general evolution equation away from the phase
transition.
\par
For a finite number of components a truncation to a finite number of couplings
becomes necessary and the evolution equation has to be solved numerically.
We have concentrated on
the phase transition of the
three-dimensional theory. Our results for the critical exponents
$\beta$, $\nu$, $\gamma$, $\delta$ agree at the
1-5 \% level with the results of the most sophisticated calculations
through other methods (summed perturbation series in three dimensions,
$\ex$-expansion, lattice calculations, $1/N$-expansion).
The anomalous dimension $\eta$ is also well determined, even though
significant improvement is expected with a more efficient treatment.
We also emphasize that, apart from the precise
determination of the critical dynamics,
the method of the effective average action permits the
discussion of the non-universal behaviour of the theory, such as the
behaviour away from the critical temperature for the
finite temperature four-dimensional theory \cite{transition}.
In this way permits the determination of the region of attraction
of the possible infrared fixed points.
The effective average action is not only useful for a study of
fixed points. More generally it describes the relevant physics at a
given lenght scale $k^{-1}$. In practice, this can be used through
the identification
of the infrared cutoff scale $k$
with the appropriate physical infrared cutoff
scale of the problem (such as the critical
bubble scale for first order phase transitions or the Hubble parameter for
inflationary cosmology).
\par
We conclude that the formalism of the effective average action provides
intuitive understanding, as well as computational power, for a
large number of physical problems.
Work in progress focuses on the equation of state for the
$O(N)$-symmetric theory \cite{progress}, the further
improvement of the precision for
the calculation of the critical exponents,
and the study of scalar theories with
different symmetries and possible first order phase transitions \cite{preps}.
The concept of the effective integration of degrees of freedom through
averaging has been developed for fermionic and gauge fields
\cite{fermiongauge}, and an exact evolution equation has been
formulated for gauge theories \cite{prepg}.
Work in this direction will lead to the study of more complicated systems,
with exciting possibilities for new physical behaviour.

\newpage

\setcounter{equation}{0}
\renewcommand{\theequation}{{\bf A.}\arabic{equation}}

\section*{Appendix A: The integrals $L^d_n$}

In this appendix we disuss the integrals
$L^d_n(\wt)=L^d_n(\wt,0)$,
where $L^d_n(\wt,\eta)$ are defined in eq. (\ref{fourtwo})
with $\Lt^d_n(\wt,\eta,\zh)$ given by eqs. (\ref{fourfour}).
With the choice (\ref{twotwo}) - (\ref{twothree}) for $R_k(q)$, they read
\beq
L^d_0(\wt) =~&-2 \int_0^{\infty} dy y^{\frac{d}{2}+1}
\frac{\exp(-y)}{\left[ 1-\exp(-y) \right]^2}
\left[ \frac{y}{1-\exp(-y)} + \wt \right]^{-1}
\nonumber \\
L^d_n(\wt) =~&-2 n \int_0^{\infty} dy y^{\frac{d}{2}+1}
\frac{\exp(-y)}{\left[ 1-\exp(-y) \right]^2}
\left[ \frac{y}{1-\exp(-y)} + \wt \right]^{-(n+1)}~{\rm for}~~ n
\geq 1.
\label{aone} \eeq
They obey the relations
\beq
\frac{\partial}{\partial \wt} L^d_0(\wt) =~&-L^d_1(\wt)
\nonumber \\
\frac{\partial}{\partial \wt} L^d_n(\wt) =~&-n L^d_{n+1}(\wt)
{}~~~~~~~~~~~~~~~~~~~~~~~~~~~~~~~~~~~~~~~~~~~{\rm for}~~ n \geq 1.
\label{aextra} \eeq
For $\wt > -1$, the integrals
$|L^d_n(\wt)|$ are finite, monotonically decreasing functions of $\wt$
\footnote{The pole structure of $L^d_n(\wt)$ is not relevant for this
work. For a discussion see ref. \cite{convex}.}.
We define
\be
L^d_0(0) = -2 l^d_n
\label{atwo} \ee
and give the values of $l^d_n$ for the first five values of $n$
\beq
l^d_0 =~&\zeta \left( \frac{d}{2} + 1 \right)
\Gamma \left( \frac{d}{2} + 1 \right)
\nonumber \\
l^d_1 =~&\Gamma \left( \frac{d}{2} \right)
\nonumber \\
l^d_2 =~&2 \left( 1 - 2^{1-\frac{d}{2}} \right)
\Gamma \left( \frac{d}{2} - 1 \right)
\nonumber \\
l^d_3 =~&3 \left( 1 - 2^{3-\frac{d}{2}} + 3^{2-\frac{d}{2}} \right)
\Gamma \left( \frac{d}{2} - 2 \right)
\nonumber \\
l^d_4 =~&4 \left( 1 - 3 \times 2^{3-\frac{d}{2}} + 3^{4-\frac{d}{2}}
- 4^{3-\frac{d}{2}} \right)
\Gamma \left( \frac{d}{2} - 3 \right).
\label{athree} \eeq
We define the threshold functions $s^d_n(\wt)$ through
\be
L^d_n(\wt) = -2 l^d_n s^d_n(\wt).
\label{afour} \ee
They are decreasing functions of
$\wt$, and $1 \geq s^d_n(\wt) > 0$ for $0 \leq \wt < \infty$.
The first two terms of the asymptotic expansion of $L^d_n(\wt)$
for large arguments $\wt \rightarrow \infty$ read
\beq
L^d_0(\wt) =~&-2~\zeta \left( \frac{d}{2} + 1 \right)
\Gamma \left( \frac{d}{2} + 2 \right) \wt^{-1}
+ \left[
\zeta \left( \frac{d}{2} + 1 \right)
+ \zeta \left( \frac{d}{2} + 2 \right) \right]
\Gamma \left( \frac{d}{2} + 3 \right) \wt^{-2}...
\nonumber \\
L^d_n(\wt) =~&-2 n~ \zeta \left( \frac{d}{2} + 1 \right)
\Gamma \left( \frac{d}{2} + 2 \right) \wt^{-(n+1)}
\nonumber \\
&+ n (n+1) \left[
\zeta \left( \frac{d}{2} + 1 \right)
+ \zeta \left( \frac{d}{2} + 2 \right) \right]
\Gamma \left( \frac{d}{2} + 3 \right) \wt^{-(n+2)}...~~{\rm for}~~ n \geq 1.
\label{afive} \eeq

\setcounter{equation}{0}
\renewcommand{\theequation}{{\bf B.}\arabic{equation}}

\section*{Appendix B: Wave function renormalization}

In this appendix we derive general expressions for the anomalous dimensions
$\eta$, $\etat$.
We consider an effective action of the form of eq. (\ref{twoeleven}).
This is no longer the most general ansatz contributing to
the wave function renormalization
and our equations are therefore not exact, in contrast to the ones for the
effective average potential. In addition,
we neglect the $q^2$ dependence of the couplings
$Z_k(\rho,q^2)$.
This truncation corresponds to the most general action containing
no more than two derivatives.
For notational simplicity,
we denote
in this appendix
$U_k(\rho)$, $Z_k(\rho)$, $\Zt_k(\rho)$, $Y_k(\rho)$
by $U$, $Z$, $\Zt$, $Y$ respectively.
Primes on $U$, $Z$, $\Zt$, $Y$ denote derivatives with respect to $\rho$.
The wave function renormalizations
$Z_k$, $\Zt_k$, $Y_k$ are defined at the minimum of the
potential $\rhz(k)$ according to eq.
(\ref{threeseven})
and the anomalous dimensions according to eqs. (\ref{threeeight}).
Also primes on $Z_k$, $\Zt_k$, $Y_k$ denote $\rho$-derivatives
of
$Z_k(\rho)$, $\Zt_k(\rho)$, $Y_k(\rho)$
at $\rho=\rhz$.
We point out that for $N=1$ one should set $Y=0$
and $\Zt$ is the relevant wave function renormalization.
\par
For a calculation of the evolution of $Z$ (which is connected to
the anomalous
dimension $\eta$) we need an expansion of the effective average action
$\Gamma_k$ around the configuration given by eq. (\ref{eightone}). This
calculation can be found in section 5 of ref. \cite{christof2}.
Here we present the result for the second functional derivative
$\Gamma^{(2)}_k$.
It can be written as
\be
(\Gamma^{(2)}_k)_{ab}(q,q') = (\Gamma_0)_{ab}(q,q')
+ (\Gamma_1)_{ab}(q,q') + (\Gamma_2)_{ab}(q,q'),
\label{bone} \ee
where we have separated the terms containing zero, one and two powers of
$\df$.
The first term reads
\be
(\Gamma_0)_{ab}(q,q') = \left[
\left( Z \delta_{ab} +
\rho Y \delta_{a1} \delta_{b1} \right) q^2
+ M^2_{ab}
\right] \delta(q-q'),
\label{btwo} \ee
where
\be
M^2_{ab} = U' \delta_{ab} + 2 \rho U'' \delta_{a1} \delta_{b1}.
\label{bthree} \ee
The second has only an off diagonal contribution
\beq
&(\Gamma_1)_{11}(q,q') = (\Gamma_1)_{22}(q,q') = 0,~~~~~
(\Gamma_1)_{ab}(q,q') = 0~~{\rm for}~~a,b > 2
\nonumber \\
&(\Gamma_1)_{12}(q,q') = \hat{\Gamma}_1(q,Q) \delta(q-q'-Q)
+ \hat{\Gamma}^*_1(q,-Q) \delta(q-q'+Q)
\nonumber \\
&(\Gamma_1)_{21}(q,q') = (\Gamma_1^*)_{12}(q',q)
\nonumber \\
{\rm with}~~~~~&\hat{\Gamma}_1(q,Q)= \phi \df \bigl\{ U'' + \frac{1}{2} Y q^2
-Z' (q-Q)Q \bigr\}.
\label{bfour} \eeq
The diagonal elements of the third term read
\beq
(\Gamma_2)_{aa}(q,q) =~&\df^* \df \bigl\{
U'' + 2 U'' \delta_{a2} +2 U'''\rho \delta_{a1}
\nonumber \\
&+q^2(Z'+Y \delta_{a2} +Y' \rho \delta_{a1})
+Q^2(Z'+Y \delta_{a2} + 2 Z'' \rho \delta_{a1}) \bigr\}.
\label{bfive} \eeq
Subsequently one has to evaluate the trace on the r.h.s. of eq.
(\ref{twosix}) and extract the piece proportional to
$Q^2(\df^* \df)$, which gives the evolution of $Z$ according to
eq. (\ref{eighttwo}).
The details for the calculation of the trace are given in ref.
\cite{christof2}.
For the scale dependence of $Z$ $(=Z_k(\rho))$ at fixed $\rho$
one obtains
\be
\left. \frac{\partial Z}{\partial t} \right|_\rho =
\left. \frac{\partial Z^{(a)}}{\partial t} \right|_\rho +
\left. \frac{\partial Z^{(b)}}{\partial t} \right|_\rho,
\label{bsix} \ee
with
\beq
\left. \frac{\partial Z^{(a)}}{\partial t} \right|_\rho
=~&v_d [ (N-1) Z' + Y] Z_k^{-1} k^{d-2} L^d_1(w_1)
+ v_d ( Z' + 2 Z'' \rho) Z_k^{-1} k^{d-2} \Lt^d_1(w_2)
\label{bseven} \\
\left. \frac{\partial Z^{(b)}}{\partial t} \right|_\rho
=~&4 v_d [U'']^2 \rho k^{d-6} Q^{d,0}_{2,1}(w_1,w_2)
+ 4 v_d Y  U'' \rho k^{d-4} Q^{d,1}_{2,1}(w_1,w_2)
\nonumber \\
&+ v_d Y^2 \rho k^{d-2} Q^{d,2}_{2,1}(w_1,w_2)
- 8 v_d Z' U'' \rho k^{d-4} L^{d}_{1,1}(w_1,w_2)
\nonumber \\
&- \frac{4 v_d}{d} [Z']^2 \rho k^{d-2} L^{d+2}_{1,1}(w_1,w_2)
- 4 v_d Z' Y \rho k^{d-2} L^{d+2}_{1,1}(w_1,w_2)
\nonumber \\
&+ \frac{16 v_d}{d} Z' U'' \rho k^{d-4} N^{d}_{2,1}(w_1,w_2)
+ \frac{8 v_d}{d} Z' Y \rho k^{d-2} N^{d+2}_{2,1}(w_1,w_2).
\label{beight} \eeq
Here we have used the variables
\beq
w_1 =~&U' \nonumber \\
w_2 =~&U' + 2 \rho U''.
\label{bnine} \eeq
It is also convenient to define the shorthands
\beq
P =~&Z x + R_k(x) \nonumber \\
\Pt =~&\Zt x + R_k(x)
\label{bten} \eeq
with $R_k(q)$ given by eq. (\ref{twotwo}) and $x=q^2$.
The integrals $L^d_n$, $\Lt^d_n$ are
given by
\beq
L^d_n(w) =~&k^{2n-d} Z_k^n \int_0^{\infty} dx x^{\frac{d}{2}-1}
\frac{\partial}{\partial t}
(P+w)^{-n}
\nonumber \\
\Lt^d_n(w) =~&k^{2n-d} Z_k^n \int_0^{\infty} dx x^{\frac{d}{2}-1}
\frac{\partial}{\partial t}
(\Pt+w)^{-n},
\label{beleven} \eeq
where
we emphasize that $\partial/\partial t$ acts on the
r.h.s. only on $R_k(x)$.
In the approximation of uniform wave function renormalization
(when the $\rho$ dependence of $Z$, $\Zt$ is neglected) these integrals
are discussed in the beginning of section 4 and in appendix A.
Notice the connection between the arguments $w$ and $\wt$:
$\wt = {w}/{Z_k k^2}$.
The dimensionless integrals
\be
L^d_{n_1,n_2}(w_1,w_2) =
k^{2(n_1+n_2)-d}
\int_0^{\infty} dx x^{\frac{d}{2}-1}
\frac{\partial}{\partial t}
\bigl\{ (P+w_1)^{-n_1} (\Pt+w_2)^{-n_2} \bigr\}
\label{btwelve} \ee
are generalizations of the integrals $L^d_n$, $\Lt^d_n$ with
\beq
L^d_{n,0}(w_1) =~&Z_k^{-n} L^d_n(w_1) \nonumber \\
L^d_{0,n}(w_2) =~&Z_k^{-n} \Lt^d_n(w_2).
\label{bthirteen} \eeq
Similarly, the integrals
$N^d_{n_1,n_2}(w_1,w_2)$
are defined by
\be
N^d_{n_1,n_2}(w_1,w_2)
=
k^{2(n_1+n_2-1)-d} \int_0^{\infty} dx x^{\frac{d}{2}}
\frac{\partial}{\partial t}
\bigl\{ \frac{\partial P}{\partial x}
(P+w_1)^{-n_1} (\Pt+w_2)^{-n_2} \bigr\}.
\label{bfourteen} \ee
Finally, the integrals
\beq
Q^{d,\alpha}_{n_1,n_2}&(w_1,w_2)
= k^{2(n_1+n_2-\alpha)-d} \int_0^{\infty} dx x^{\frac{d}{2}-1+\alpha}
\nonumber \\
&\frac{\partial}{\partial t}
\biggl\{
\left[
\frac{\partial P}{\partial x}
+ \frac{2x}{d} \frac{\partial^2 P}{\partial x^2}
- \frac{4x}{d} \left(\frac{\partial P}{\partial x}\right)^2
(P+w_1)^{-1}
\right]
(P+w_1)^{-n_1} (\Pt+w_2)^{-n_2} \biggr\}
\label{bfifteen} \eeq
are related by partial integration to other integrals
\be
M^d_{n_1,n_2}(w_1,w_2)
=
k^{2(n_1+n_2-1)-d} \int_0^{\infty} dx x^{\frac{d}{2}}
\frac{\partial}{\partial t}
\biggl\{ \left( \frac{\partial P}{\partial x} \right)^2
(P+w_1)^{-n_1} (\Pt+w_2)^{-n_2} \biggr\}
\label{bsixteen} \ee
through
\beq
Q^{d,\alpha}_{n_1,n_2}(w_1,w_2)
=~&\frac{2 n_1-4}{d} M^{d+2 \alpha}_{n_1+1,n_2}(w_1,w_2)
+ \frac{2 n_2}{d} M^{d+2 \alpha}_{n_1,n_2+1}(w_1,w_2)
\nonumber \\
&+ \frac{2 n_2}{d} \rho Y N^{d+2 \alpha}_{n_1,n_2+1}(w_1,w_2)
- \frac{2 \alpha}{d} N^{d+2 \alpha -2}_{n_1,n_2}(w_1,w_2).
\label{bseventeen} \eeq
The evolution equation for $Z$ depends, in this approximation,
on $U'$, $U''$, $Z$, $Z'$, $Z''$ and $Y$.
There is also an $\eta$ dependence resulting from the $t$-derivative
acting on $Z_k$ in $R_k(q)$.
\par
We define the anomalous dimension $\eta$ by the variation of
$\ln Z_k = \ln Z_k(\rhz(k))$ (see eqs. (\ref{threeseven}),
(\ref{threeeight})). In the \underline{spontaneously broken regime},
where $\rhz(k) \not= 0$, this implies
\be
\eta = - Z_k^{-1}
\left(
\left. \frac{\partial Z^{(a)}}{\partial t} \right|_{\rhz} +
\left. \frac{\partial Z^{(b)}}{\partial t} \right|_{\rhz}
\right)
- Z_k^{-1} Z'_k \db.
\label{beighteen} \ee
Here $\db=d \rhz(k)/dt$ describes the k dependence of the minimum of the
potential
\beq
\db =~&-v_d Z_k^{-1} k^{d-2}
\bigl\{ (N-1) L^d_1(0) + \left( 3 + \frac{2 U_3 \rhz}{\lb} \right)
\Lt^d_1(2 \lb \rhz) \bigr\}
\nonumber \\
&-v_d Z_k^{-1} k^{d}
\bigl\{
(N-1) \frac{Z'_k}{\lb} L^{d+2}_1(0)
+ \frac{\Zt'_k}{\lb} \Lt^{d+2}_1(2 \lb \rhz)
\bigr\},
\label{bnineteen} \eeq
with
\beq
&\lb =U''_k(\rhz),~~~~~U_n = U^{(n)}_k(\rhz)~~~{\rm for}~~n \geq 3
\nonumber \\
&w_1(\rhz) =0,~~~~~~w_2(\rhz) = 2 \lb \rhz.
\label{btwenty} \eeq
Combining eqs. (\ref{bseven}), (\ref{beight}), (\ref{beighteen}),
(\ref{bnineteen}) we obtain
\beq
\eta =~&-v_d k^{d-2} Z_k^{-2}
\bigl\{ \left[ (N-1)Z'_k + Y_k \right] L^d_1(0)
+ (Z'_k + 2 \rhz Z''_k) \Lt^d_1(2 \lb \rhz) \bigr\}
\nonumber \\
&- \frac{8 v_d}{d} Z_k^{-1} \lb^2 \rhz k^{d-6}
\biggl\{
M^d_{2,2}(0,2 \lb \rhz) + \frac{Y_k k^2}{\lb} M^{d+2}_{2,2}(0,2 \lb \rhz)
+ \frac{1}{4}
\left( \frac{Y_k k^2}{\lb} \right)^2 M^{d+4}_{2,2}(0,2 \lb \rhz)
\biggr\}
\nonumber \\
&- \frac{8 v_d}{d} Z_k^{-1} Y_k \lb^2 \rhz^2 k^{d-6}
\biggl\{
N^d_{2,2}(0,2 \lb \rhz) + \frac{Y_k k^2}{\lb} N^{d+2}_{2,2}(0,2 \lb \rhz)
+ \frac{1}{4}
\left( \frac{Y_k k^2}{\lb} \right)^2 N^{d+4}_{2,2}(0,2 \lb \rhz)
\biggr\}
\nonumber \\
&- \frac{8 v_d}{d} Z_k^{-1} (2 Z'_k - Y_k)\lb \rhz k^{d-4}
\biggl\{
N^d_{2,1}(0,2 \lb \rhz) +
\frac{1}{2} \frac{Y_k k^2}{\lb} N^{d+2}_{2,1}(0,2 \lb \rhz)
\biggr\}
\nonumber \\
&+8 v_d Z_k^{-1} Z'_k \lb \rhz k^{d-4} L^d_{1,1}(0,2 \lb \rhz)
+ \frac{4 v_d}{d} Z_k^{-1} Z'_k (Z'_k + d Y_k)
\rhz k^{d-2} L^{d+2}_{1,1}(0,2 \lb \rhz)
\nonumber \\
&-Z_k^{-1} Z'_k \db.
\label{btwentyone} \eeq
It is convenient to bring $\eta$ into the form
\beq
\eta=~&- \frac{2 v_d}{d} Z_k^{-1} \rhz^{-1} k^{d-2}
\bigl\{ M^d_{2,0}(0) - 2 M^d_{1,1}(0,2 \lb \rhz) + M^d_{0,2}(2 \lb \rhz)
\bigr\}
\nonumber \\
&-v_d Z_k^{-2} Y_k k^{d-2}
\bigl\{ \Lt^d_1(2 \lb \rhz) -\frac{2}{d} Z_k^{-1} Y_k \rhz
\Lt^{d+2}_2(2 \lb \rhz) \bigr\}
\nonumber \\
&-v_d Z_k^{-1} Z'_k k^{d-2}
\biggl\{ (N-1) Z_k^{-1} L^d_1(0) - \frac{8}{d} N^d_{1,1}(0,2 \lb \rhz)
\nonumber \\
&+ Z_k^{-1} \left( 5 + \frac{2 Z''_k \rhz}{Z'_k} \right) \Lt^d_1(2 \lb \rhz)
- \frac{4}{d} Z'_k \rhz L^{d+2}_{1,1}(0,2 \lb \rhz)
\biggr\}
\nonumber \\
&-Z_k^{-1} Z'_k \db.
\label{btwentytwo} \eeq
For this purpose we have employed the following identities
\beq
N^{d}_{n_1,n_2}&(w_1,w_2) =
\nonumber \\
&\frac{k^2}{w_2-w_1}
\bigl\{ N^{d}_{n_1,n_2-1}(w_1,w_2)
- N^{d}_{n_1-1,n_2}(w_1,w_2)
-Y \rho N^{d+2}_{n_1,n_2}(w_1,w_2)
\bigr\}
\label{btwentythree} \\
N^{d}_{n,0}&(0) = \frac{d}{2(n-1)} Z_k^{-(n-1)}
L^d_{n-1}(0)~~~~~~~~~~~~~~~~~~~~~~~~~~~{\rm for}~~n>1
\label{btwentyfour} \\
N^{d}_{0,n}&(w) =
\frac{d}{2(n-1)} Z_k^{-(n-1)} \Lt^d_{n-1}(w)
- Z_k^{-n} Y \rho \Lt^{d+2}_{n}(w)~~~{\rm for}~~n>1
\label{btwentyfive} \\
M^{d}_{n_1,n_2}&(w_1,w_2) =
\nonumber \\
&\frac{k^2}{w_2-w_1}
\bigl\{ M^{d}_{n_1,n_2-1}(w_1,w_2)
-M^{d}_{n_1-1,n_2}(w_1,w_2)
-Y \rho M^{d+2}_{n_1,n_2}(w_1,w_2)
\bigr\}.
\label{btwentysix} \eeq
We also define the function
\beq
\Mh^d_2(w) =~&M^d_{2,0}(0)
- 2 M^d_{1,1}(0,w) + M^d_{0,2}(w)
\nonumber \\
=~&k^{2-d} \int_0^{\infty} dx x^{\frac{d}{2}}
\frac{\partial }{ \partial t} \biggl\{
\frac{\partial P}{ \partial x}
\left( \frac{1}{P} - \frac{1}{\Pt + w} \right)
\biggr\}^2.
\label{btwentyseven} \eeq
\par
For the calculation of the evolution of $\Zt$ (which is connected to
$\etat$) we need an expansion of $\Gamma_k$ around the configuration given
by eq. (\ref{eightthree}).
The second functional derivative
$\Gamma^{(2)}_k$ is again given by eq. (\ref{bone}) with
$\Gamma_0$ given by eqs. (\ref{btwo}), (\ref{bthree}).
The other two terms now read
\beq
(\Gamma_1)_{ab}&(q,q') = (\hat{\Gamma}_1)_{ab}(q,Q) \delta(q-q'-Q)
+ (\hat{\Gamma}^*_1)_{ab}(q,-Q) \delta(q-q'+Q)
\nonumber \\
(\Gamma_1)_{ba}&(q,q') = (\Gamma_1^*)_{ab}(q',q)
\nonumber \\
{\rm with}~~(\hat{\Gamma}_1)_{ab}&(q,Q) = \phi \df
\biggl\{
\left[ U'' +(q^2-qQ) Z' + \frac{1}{2} Q^2 Y \right] \delta_{ab}
\nonumber \\
&+ \left[ 2 U'' +2 U''' \rho + (q^2-qQ)(Y + Y' \rho)
+Q^2(Z' + \frac{1}{2} Y + Y' \rho) \right] \delta_{a1} \delta_{b1}
\biggr\},~~~~~~
\label{btwentyeight} \eeq
and
\beq
(\Gamma_2)_{ab}(q,q) =~&\df^* \df \bigl\{
(U'' + 2 U''' \rho) \delta_{ab}
+ (2 U'' + 10 U''' \rho +4 U^{(4)} \rho^2) \delta_{a1} \delta_{b1}
\nonumber \\
&+(Z'+Y' \rho)Q^2 \delta_{ab} + (Y + 2 Z'' \rho +4 Y' \rho
+ 2 Y'' \rho^2) Q^2 \delta_{a1} \delta_{b1}
\nonumber \\
&+(Z' + 2 Z'' \rho)q^2 \delta_{ab}
+ (Y+ 5Y' \rho +2 Y'' \rho^2) q^2 \delta_{a1} \delta_{b1}
\bigr\}.
\label{btwentynine} \eeq
For the calculation of the trace we refer the reader again to ref.
\cite{christof2}
and we give the result for
\be
\left. \frac{\partial \Zt}{\partial t} \right|_\rho =
\left. \frac{\partial \Zt^{(a)}}{\partial t} \right|_\rho +
\left. \frac{\partial \Zt^{(b)}}{\partial t} \right|_\rho.
\label{bthirty} \ee
One finds (with $w_1$, $w_2$ given by eqs. (\ref{bnine}))
\beq
\left. \frac{\partial \Zt^{(a)}}{\partial t} \right|_\rho&
= v_d  (N-1) (Z' + Y' \rho) Z_k^{-1} k^{d-2} L^d_1(w_1)
\nonumber \\
&+ v_d ( Z' + 2 Z'' \rho + Y + 5 Y' \rho + 2 Y'' \rho^2)
Z_k^{-1} k^{d-2} \Lt^d_1(w_2)
\label{bthirtyone} \\
\left. \frac{\partial \Zt^{(b)}}{\partial t}\right|_\rho&
= 2 v_d (N-1) [U'']^2 \rho k^{d-6} Q^{d,0}_{3,0}(w_1)
+ 4 v_d (N-1) U'' Z' \rho k^{d-4} Q^{d,1}_{3,0}(w_1)
\nonumber \\
&+ 2 v_d (N-1) [Z']^2 \rho k^{d-2} Q^{d,2}_{3,0}(w_1)
+ 2 v_d (3 U'' + 2 U''' \rho)^2 \rho k^{d-6} \Qt^{d,0}_{3,0}(w_2)
\nonumber \\
&+ 4 v_d (Z' +Y + Y' \rho)(3 U'' + 2 U''' \rho)
\rho k^{d-4} \Qt^{d,1}_{3,0}(w_2)
+ 2 v_d (Z' +Y + Y' \rho)^2
\rho k^{d-2} \Qt^{d,2}_{3,0}(w_2)
\nonumber \\
&+ \frac{8 v_d}{d} (N-1) Z' U''
\rho k^{d-4} N^{d}_{3,0}(w_1)
+ \frac{8 v_d}{d} (N-1) [Z']^2
\rho k^{d-2} N^{d+2}_{3,0}(w_1)
\nonumber \\
&+ \frac{8 v_d}{d} (Z' +Y + Y' \rho)(3 U'' + 2 U''' \rho)
\rho k^{d-4} \Nt^{d}_{3,0}(w_2)
+ \frac{8 v_d}{d} (Z' +Y + Y' \rho)^2
\rho k^{d-2} \Nt^{d+2}_{3,0}(w_2)
\nonumber \\
&- 2 v_d (N-1) Y  U'' Z_k^{-2}
\rho k^{d-4} L^{d}_{2}(w_1)
- 2 v_d (N-1)
\left( Z' Y + \frac{1}{d} [Z']^2 \right) Z_k^{-2}
\rho k^{d-2} L^{d+2}_{2}(w_1)
\nonumber \\
&- 4 v_d (Z' +Y + Y' \rho)(3 U'' + 2 U''' \rho) Z_k^{-2}
\rho k^{d-4} \Lt^{d}_{2}(w_2)
\nonumber \\
&- 2 \left( 2 + \frac{1}{d} \right) v_d
(Z' +Y + Y' \rho)^2 Z_k^{-2}
\rho k^{d-2} \Lt^{d+2}_{2}(w_2).
\label{bthirtytwo} \eeq
Here we use the definitions
\be
\Nt^d_{n_1,n_2}(w_1,w_2) =
N^d_{n_1,n_2}(w_1,w_2; P \leftrightarrow \Pt) =
N^d_{n_2,n_1}(w_2,w_1) + Y \rho
L^{d+2}_{n_2,n_1}(w_2,w_1)
\label{bthirtythree} \ee
and
\be
\Qt^{d, \alpha}_{n_1,n_2}(w_1,w_2) =
Q^{d, \alpha}_{n_1,n_2}(w_1,w_2; P \leftrightarrow \Pt).
\label{bthirtyfour} \ee
The integrals $\Qt^{d, \alpha}_{n_1,n_2}$
are related to $\Mt^{d}_{n_1,n_2}$
by an equation analogous to eq. (\ref{bseventeen}). The relation between
$\Mt^{d}_{n_1,n_2}$ and $M^{d}_{n_1,n_2}$
reads
\beq
\Mt^{d}_{n_1,n_2}(w_1,w_2) =~&M^{d}_{n_1,n_2}(w_1,w_2; P \leftrightarrow \Pt) =
\nonumber \\
&M^{d}_{n_2,n_1}(w_2,w_1)
+ 2 Y \rho N^d_{n_2,n_1}(w_2,w_1) + Y^2 \rho^2
L^{d+2}_{n_2,n_1}(w_2,w_1).
\label{bthirtyfive} \eeq
We also observe a general relation for the integrals
$N^{d}_{n_1,n_2}(w_1,w_2)$
\beq
(n_1-1)&N^{d}_{n_1,n_2}(w_1,w_2)
= \frac{d}{2} L^{d}_{n_1-1,n_2}(w_1,w_2)
- n_2 \Nt^{d}_{n_2+1,n_1-1}(w_2,w_1)=
\nonumber \\
&\frac{d}{2} L^{d}_{n_1-1,n_2}(w_1,w_2)
- n_2 Y \rho L^{d+2}_{n_1-1,n_2+1}(w_1,w_2)
- n_2 N^{d}_{n_1-1,n_2+1}(w_1,w_2).
\label{bthirysix} \eeq
{}From this, or directly from partial integration of
eq. (\ref{bfourteen}), follow the identities
\beq
N^d_{3,0}(w) =~&\frac{d}{4} L^d_{2,0}(w) \nonumber \\
\Nt^d_{3,0}(w) =~&\frac{d}{4} L^d_{0,2}(w) \nonumber \\
N^d_{0,3}(w) =~&\frac{d}{4} L^d_{0,2}(w)
- Y \rho L^{d+2}_{0,3}(w) \nonumber \\
\Nt^d_{0,3}(w) =~&\frac{d}{4} L^d_{2,0}(w)
+ Y \rho L^{d+2}_{3,0}(w).
\label{bthirtyseven} \eeq
Using the various relations among the integrals we can express
$\partial \Zt/ \partial t$ in terms of
$M^d_{4,0}$, $\Mt^d_{4,0}$, $L^d_n$ and $\Lt^d_n$
only.
We conclude that the evolution equation for $\Zt$ $(=\Zt_k(\rho))$
depends on $U'$, $U''$, $U'''$, $Z$, $Z'$, $Z''$, $Y$, $Y'$ and
$Y''$ in our approximation.
There is also an $\eta$ dependence resulting from the $t$-derivative
acting on $Z_k$ in $R_k(q)$.
\par
The anomalous dimension for the radial mode is defined in the
\underline{spontaneously broken} \underline{regime} by
\be
\etat = - \Zt_k^{-1}
\left(
\left. \frac{\partial \Zt^{(a)}}{\partial t} \right|_{\rhz} +
\left. \frac{\partial \Zt^{(b)}}{\partial t} \right|_{\rhz}
\right)
- \Zt_k^{-1} \Zt'_k \db.
\label{bthirtyeight} \ee
We find
\beq
\etat =&-v_d Z_k^{-1} \Zt_k^{-1} k^{d-2}
\bigl\{
(N-1) (Z'_k+ Y'_k \rhz) L^d_1(0) +
(Z'_k+ 2 Z''_k \rhz +Y_k + 5 Y'_k \rhz + 2 Y''_k \rhz^2) \Lt^d_1( 2 \lb \rhz)
\bigr\}
\nonumber \\
&- \frac{4 v_d}{d} \Zt_k^{-1} \rhz k^{d-6}
\bigl\{
(N-1) \lb^2 M^d_{4,0}(0)
+ 2 (N-1) Z'_k \lb k^2 M^{d+2}_{4,0}(0)
+ (N-1) [Z'_k]^2 k^4 M^{d+4}_{4,0}(0)
\nonumber \\
&+(3 \lb + 2 U_3 \rhz)^2 \Mt^d_{4,0}( 2 \lb \rhz)
+ 2 (Z'_k + Y_k + Y'_k \rhz)(3 \lb + 2 U_3 \rhz) k^2
\Mt^{d+2}_{4,0}( 2 \lb \rhz)
\nonumber \\
&+ (Z'_k + Y_k + Y'_k \rhz)^2 k^4 \Mt^{d+4}_{4,0}( 2 \lb \rhz)
\bigr\}
\nonumber \\
&+ 2 v_d Z_k^{-2} \Zt_k^{-1} \rhz k^{d-4}
\biggl\{
(N-1) Y_k \lb L^d_2(0)
+ (N-1) \left( Z'_k Y_k + \frac{1}{d} [Z'_k]^2 \right)
k^2 L^{d+2}_2(0)
\nonumber \\
&+ 2 (Z'_k + Y_k + Y'_k \rhz)(3 \lb + 2 U_3 \rhz)
\Lt^d_2( 2 \lb \rhz)
+ \left( 2 + \frac{1}{d} \right)
(Z'_k + Y_k + Y'_k \rhz)^2 k^2 \Lt^{d+2}_2( 2 \lb \rhz)
\biggr\}
\nonumber \\
&- \Zt_k^{-1} (Z'_k + Y_k + Y'_k \rhz) \db.
\label{bthirtynine} \eeq
\par
Finally, we need the anomalous dimensions in the
\underline{symmetric regime} where $\rhz(k)=0$.
In this regime $Z_k$ and $\Zt_k$
coincide. The contribution from $\partial Z^{(b)} / \partial t$
or $\partial \Zt^{(b)} / \partial t$
vanishes identically for $\rhz=0$, since there is no cubic vertex.
One finds from eq. (\ref{bseven}) or eq. (\ref{bthirtyone})
\be
\eta = \etat =- \frac{d \ln Z_k(0)}{d t} =
- \frac{d \ln \Zt_k(0)}{d t} =
-v_d Z_k^{-2} (N Z'_k(0) + Y_k(0)) k^{d-2}
L^d_1(\mb)
\label{bfourty} \ee
with
$\mb = U'_k(0)$.

\setcounter{equation}{0}
\renewcommand{\theequation}{{\bf C.}\arabic{equation}}

\section*{Appendix C: The integral $\Mh^d_2$}

In this appendix we disuss the integral
$\Mh^d_2(\wt)=\Mh^d_2(\wt,0,1)$,
where $\Mh^d_2(\wt,\eta,\zh)$ is defined in eq. (\ref{eighteleven}).
Since $\zh=1$, it can be written as
\be
\Mh^d_2(\wt) = \wt^2 Z_k^2 M^d_{2,2}(0,\wt),
\label{cone} \ee
where $M^d_{n_1,n_2}(w_1,w_2)$ is defined in eq. (\ref{bsixteen}).
With the above approximations (uniform wave function renormalization,
$\eta=0$, $\zh=1$) and the
choice (\ref{twotwo}) - (\ref{twothree}) for $R_k(q)$,
this integral reads
\beq
M^d_{2,2}(0,\wt) =&-2 Z_k^{-2} \int_0^{\infty} dy y^{\frac{d}{2}-2}
\frac{1 + r +y \frac{\partial r}{ \partial y}}{(1+r)^2
\left[(1+r)y +\wt \right]^2 }
\nonumber \\
&\biggl\{
2 y \frac{\partial r}{ \partial y}
+ 2 \left(y \frac{\partial }{ \partial y} \right)^2 r
- 2 y^2 \left( 1 +r + y \frac{\partial r}{ \partial y} \right)
\frac{\partial r}{ \partial y}
\left[ \frac{1}{(1+r)y} + \frac{1}{(1+r)y +\wt} \right]
\biggr\},
\nonumber \\
{}~&~
\label{ctwo} \eeq
with
\be
r(y) = \frac{\exp(-y)}{1 - \exp(-y)}.
\label{cthree} \ee
The function $M^d_{2,2}(0,\wt)$ is monotonically decreasing for
positive $\wt$.
We define
\be
M^d_{2,2}(0,0) = -2 Z_k^{-2} m^d_4
\label{cfour} \ee
and find
\be
m^d_4 = 2 \Gamma \left( \frac{d}{2} -1 \right)
\left[ 1 - 2^{ - \frac{d}{2}} \left( 1 + \frac{d}{2} \right) \right].
\label{cfive} \ee
We can define the function $t^d_{2,2}(\wt)$ through
\be
M^d_{2,2}(0,\wt) = -2 Z_k^{-2} m^d_4 t^d_{2,2}(\wt).
\label{csix} \ee
It is a decreasing function of
$\wt$, and $1 \geq t^d_{2,2}(\wt) > 0$ for $0 \leq \wt < \infty$.
The first term of the asymptotic expansion of
$M^d_{2,2}(0,\wt)$ for $\wt \rightarrow \infty$ reads
\be
M^d_{2,2}(0,\wt) = 4 Z_k^{-2} \Gamma \left( \frac{d}{2} \right)
\biggl\{
\left[ 1 -d + \frac{d(d+2)}{8} \right] \zeta \left( \frac{d}{2} \right)
+ \frac{(2-d)d}{8} \zeta \left( \frac{d}{2} +1 \right)
\biggr\}
\wt^{-2}...
\label{cseven} \ee
For $d=2$ the leading term takes the particularly simple form
\be
M^2_{2,2}(0,\wt) = -2 Z_k^{-2} \wt^{-2}...
\label{ceight} \ee
and
\be
\lim_{\wt \rightarrow \infty} \Mh^2_2(\wt) = -2.
\label{cnine} \ee

\newpage

\section*{Tables}

\begin{table} [h]
\renewcommand{\arraystretch}{1.5}
\hspace*{\fill}
\begin{tabular}{|c|c|c|c|c|c|c|}	\hline
$\kx_*$
& $\lx_*$
& $u_{3*}$
& $\eta_*$
& $u_{4*}$
& $u_{5*}$
& $u_{6*}$
\\ \hline \hline
$6.57 \times 10^{-2}$
& 11.5
&
&
&
&
&
\\ \hline
$8.01 \times 10^{-2}$
& 7.27
& 52.8
&
&
&
&
\\ \hline
$7.86 \times 10^{-2}$
& 6.64
& 42.0
& $3.58 \times 10^{-2}$
&
&
&
\\ \hline
$7.75 \times 10^{-2}$
& 6.94
& 43.5
& $3.77 \times 10^{-2}$
& 95.7
&
&
\\ \hline
$7.71 \times 10^{-2}$
& 7.03
& 43.4
& $3.83 \times 10^{-2}$
& 111
& $-1.43 \times 10^3$
& $3.72 \times 10^4$
\\ \hline
\end{tabular}
\hspace*{\fill}
\renewcommand{\arraystretch}{1}
\caption[y]
{
Fixed point values for the minimum of $u$, its derivatives and the anomalous
dimension, for various truncations of the infinite system
of evolution equations. $N=3$.
}
\end{table}

\begin{table} [h]
\renewcommand{\arraystretch}{1.5}
\hspace*{\fill}
\begin{tabular}{|c|c|c|c|c|c|c|}	\hline
$N$
& $\kx_*$
& $\lx_*$
& $u_{3*}$
& $u_{4*}$
& $\eta_*$
& $2 \kx_* \lx_*$
\\ \hline \hline
1
& $4.05 \times 10^{-2}$
& 9.25
& 87.4
& 598
&$4.49 \times 10^{-2}$
& 0.749
\\ \hline
2
& $5.81 \times 10^{-2}$
& 8.08
& 61.5
& 249
&$4.23 \times 10^{-2}$
& 0.939
\\ \hline
3
& $7.71 \times 10^{-2}$
& 7.03
& 43.4
& 111
&$3.83 \times 10^{-2}$
& 1.08
\\ \hline
4
& $9.71 \times 10^{-2}$
& 6.15
& 31.3
& 56.6
&$3.42 \times 10^{-2}$
& 1.20
\\ \hline
10
& $2.26 \times 10^{-1}$
& 3.26
& 7.60
& 6.01
&$1.87 \times 10^{-2}$
& 1.48
\\ \hline
20
& $4.49 \times 10^{-1}$
& 1.77
& 2.14
& 0.944
& $1.02 \times 10^{-2}$
& 1.59
\\ \hline
100
& 2.24
& 0.375
& $9.32 \times 10^{-2}$
& $9.05 \times 10^{-3}$
& $2.17 \times 10^{-3}$
& 1.68
\\ \hline
\end{tabular}
\hspace*{\fill}
\renewcommand{\arraystretch}{1}
\caption[y]
{
Fixed point values for the minimum of $u$, its derivatives and the anomalous
dimension, for various values of $N$, with the truncation II (eq.
(\ref{seveneleven})).
}
\end{table}

\newpage

\begin{table} [h]
\renewcommand{\arraystretch}{1.5}
\hspace*{\fill}
\begin{tabular}{|c||c|c||c|c||c|c||c|c|}	\hline
$N$
&\multicolumn{2}{c||}{$\beta$}
&\multicolumn{2}{c||}{$\nu$}
&\multicolumn{2}{c||}{$\gamma$}
&\multicolumn{2}{c|}{$\eta_*$}
\\
\hline \hline

&
&$0.3250(15)^a$
&
&$0.6300(15)^a$
&
&$1.2405(15)^{a}$
&
&$0.032(3)^{a}$
\\
1
&0.333
&$0.3270(15)^{b}$
&0.638
&$0.6310(15)^{b}$
&1.247
&$1.2390(25)^{b}$
&0.045
&$0.0375(25)^{b}$
\\

&
&$0.312(5)^{c}$
&
&$0.6305(15)^{c}$
&
&$1.2385(25)^{c}$
&
&
\\ \hline

&
&$0.3455(20)^{a}$
&
&$0.6695(20)^{a}$
&
&$1.316(25)^{a}$
&
&$0.033(4)^{a}$
\\
2
&0.365
&$0.3485(35)^{b}$
&0.700
&$0.671(5)^{b}$
&1.371
&$1.315(7)^{b}$
&0.042
&$0.040(3)^{b}$
\\

&
&
&
&$0.672(7)^{c}$
&
&$1.33(2)^{c}$
&
&
\\ \hline

&
&$0.3645(25)^{a}$
&
&$0.705(3)^{a}$
&
&$1.386(4)^{a}$
&
&$0.033(4)^{a}$
\\
3
&0.390
&$0.368(4)^{b}$
&0.752
&$0.710(7)^{b}$
&1.474
&$1.39(1)^{b}$
&0.038
&$0.040(3)^{b}$
\\

&
&$0.38(3)^{c}$
&
&$0.715(20)^{c}$
&
&$1.38(2)^{c}$
&
&
\\ \hline
4
&0.409
&
&0.791
&
&1.556
&
&0.034
&
\\  \hline
10
&0.461
&$0.449^{d}$
&0.906
&$0.877^{d}$
&1.795
&$1.732^{d}$
&0.019
&$0.025^{d}$
\\ \hline
20
&0.481
&$0.477^{d}$
&0.952
&$0.942^{d}$
&1.895
&$1.872^{d}$
&0.010
&$0.013^{d}$
\\ \hline
100
&0.497
&$0.496^{d}$
&0.992
&$0.989^{d}$
&1.981
&$1.975^{d}$
&0.002
&$0.003^{d}$
\\ \hline
\end{tabular}
\hspace*{\fill}
\renewcommand{\arraystretch}{1}
\caption[y]
{
Critical exponents of the three-dimensional theory for various values of $N$.
For comparison we list results obtained with other methods as summarized in
\cite{zinn} and \cite{journal}: \\
a) From summed perturbation series in fixed dimension 3 at six-loop order. \\
b) From the $\ex$-expansion at order $\ex^5$. \\
c) From lattice calculations. \\
d) From the $1/N$-expansion at order $1/N^2$.
}
\end{table}

\begin{table} [h]
\renewcommand{\arraystretch}{1.5}
\hspace*{\fill}
\begin{tabular}{|c|||c|c||c|c||c|c||c|c||c|c||c|c||c|c|}	\hline
$N$
&\multicolumn{2}{c||}{1}
&\multicolumn{2}{c||}{2}
&\multicolumn{2}{c||}{3}
&\multicolumn{2}{c||}{4}
&\multicolumn{2}{c||}{10}
&\multicolumn{2}{c||}{20}
&\multicolumn{2}{c|}{100}
\\
\hline
$\frac{\lx_R}{m_R}$
& 9.6
& $7.9^a$
& 8.4
& $7.1^a$
& 7.4
& $6.4^a$
& 6.6
& ~~~~
& 3.8
& $5.0^b$
& 2.1
& $2.5^b$
& 0.49
& $0.50^b$
\\ \hline
\end{tabular}
\hspace*{\fill}
\renewcommand{\arraystretch}{1}
\caption[y]
{
The ratio ${\lx_R}/{m_R}$ for
the three-dimensional theory in the critical region, for various values of $N$.
For comparison we list results obtained with other methods: \\
a) From summed perturbation series in fixed dimension 3 \cite{parisi,zinn}. \\
b) From large $N$ calculations \cite{largen,zinn}.
}
\end{table}

\newpage
\section*{Figures}

\renewcommand{\labelenumi}{Fig. \arabic{enumi}}
\begin{enumerate}
\item  
Graphical representation of the exact evolution equation for the
effective average action $\Gamma_k$.
\item  
Solution of the evolution equations for $\kx(k)$, $\lx(k)$, $u_3(k)$ and
$\eta(k)$ in the critical region, for initial values slightly above and below
the critical line.
The fixed point for the second order phase transition is displayed, as well as
the final running towards
the phase with spontaneous symmetry breaking or the
symmetric one. $N =3$.
\item  
The critical exponents as the phase transition is appoached. $N=4$.
\end{enumerate}


\newpage

\begin{thebibliography}{99}


\bibitem{exact}
C. Wetterich, Phys. Lett. B {\bf 301}, 90 (1993).

\bibitem{kadanoff}
L.P. Kadanoff, Physics {\bf 2}, 263 (1966).

\bibitem{wilson}
K.G. Wilson, Phys. Rev. B {\bf 4}, 3174;3184 (1971);
K.G. Wilson and I.G. Kogut, Phys. Rep. {\bf 12}, 75 (1974);
F.J. Wegner, in: Phase Transitions and Critical Phenomena, vol. 6,
eds. C. Domb and M.S. Greene, Academic Press (1976).

\bibitem{schwinger}
F.J. Dyson, Phys. Rev. {\bf 75}, 1736 (1949);
J. Schwinger, Proc. Nat. Acad. Sc. {\bf 37}, 452;455 (1951).

\bibitem{polchinski}
J. Polchinski, Nucl. Phys. B {\bf 231}, 269 (1984);

\bibitem{private}
U. Ellwanger, private communication;
M. Bonini, M. D'Attanasio and G. Marchesini, Parma preprint UPRF-92-360.

\bibitem{christof1}
C. Wetterich, Nucl. Phys. B {\bf 352}, 529 (1991).

\bibitem{more}
C. Wetterich, Heidelberg preprint HD-THEP-92-64, in press Z. Phys. C.

\bibitem{mmore}
C. Wetterich, Heidelberg preprint HD-THEP-93-17.

\bibitem{ellwanger}
U. Ellwanger, Heidelberg preprint HD-THEP-92-33.

\bibitem{calsym}
C.G. Callan, Phys. Rev. D {\bf 2}, 1541 (1970);
K. Symanzik, Commun. Math. Phys. {\bf 18}, 227 (1970).

\bibitem{zinn}
J. Zinn-Justin, Quantum field theory and critical phenomena, Oxford Science
Publications (1989);

\bibitem{parisi}
G. Parisi, J. Stat. Phys. {\bf23}, 49 (1980); Statistical field theory,
Addison-Wesley (1988).

\bibitem{christof2}
C. Wetterich, Z. Phys. C {\bf 57}, 451 (1993).

\bibitem{transition}
N. Tetradis and C. Wetterich, Nucl. Phys. B. {\bf 398}, 659 (1993).

\bibitem{jain}
V. Jain, Nucl. Phys. B {\bf 394}, 707 (1993);
H. Meyer-Ortmanns and A. Patkos, Phys. Lett. B {\bf 297}, 331 (1992).

\bibitem{largen}
M. Reuter, N. Tetradis and C. Wetterich, preprint DESY-93-004,
in press Nucl. Phys. B.

\bibitem{convex}
A. Ringwald and C. Wetterich, Nucl. Phys. B {\bf 334}, 506 (1990);
N. Tetradis and C. Wetterich, Nucl. Phys. B {\bf 383}, 197 (1992).

\bibitem{progress}
N. Tetradis and C. Wetterich, in preparation.

\bibitem{largenn}
H.J. Schnitzer, Phys. Rev. D {\bf 10}, 1800 (1974);
S. Coleman, R. Jackiw and H.D. Politzer, Phys. Rev. D {\bf 10}, 2491 (1974);
L.F. Abbott, J.S. Kang and H.J. Schnitzer, Phys. Rev. D {\bf 13}, 2212 (1976);
W.A. Bardeen and M. Moshe, Phys. Rev. D {\bf 28}, 1372 (1983).

\bibitem{journal}
I. Kondor and T. Temesvari, J. Phys. Lett. (Paris) {\bf 39}, L99 (1978).

\bibitem{preps}
S. Bornholdt, N. Tetradis and C. Wetterich, in preparation.

\bibitem{fermiongauge}
C. Wetterich, Z. Phys. C {\bf 48}, 693 (1990);
M. Reuter and C. Wetterich, Nucl. Phys. B {\bf 391}, 147 (1993);
Heidelberg preprint HD-THEP-92-62;
S. Bornholdt and C. Wetterich, Phys. Lett. B {\bf 282}, 399 (1992).

\bibitem{prepg}
M. Reuter and C. Wetterich, in preparation.


\end{thebibliography}
\end{document}